\documentclass[showpacs,amsmath,amssymb,amsfonts,twocolumn,floatfix,superscriptaddress]{revtex4} %superscriptaddress %%%,twocolumn
\usepackage{graphicx}
\usepackage{bm}
\usepackage[german,english]{babel}

\begin{document}

\title{Two-component few-fermion mixtures in a one-dimensional trap: numerical versus analytical approach}

\author{Ioannis Brouzos}
\email{ibrouzos@physnet.uni-hamburg.de}
\affiliation{Zentrum f\"ur Optische Quantentechnologien, Universit\"at Hamburg, Luruper Chaussee 149, 22761 Hamburg, Germany}

\author{Peter Schmelcher}
\email{pschmelc@physnet.uni-hamburg.de}
\affiliation{Zentrum f\"ur Optische Quantentechnologien, Universit\"at Hamburg, Luruper Chaussee 149, 22761 Hamburg, Germany}


\begin{abstract}
We explore a few-fermion mixture consisting of two components which are repulsively interacting and confined in a one-dimensional harmonic trap. Different scenarios of population imbalance ranging from the completely imbalanced case where the physics of a single impurity in the Fermi-sea is discussed to the partially imbalanced and equal population configurations are investigated. For the numerical calculations the multi-configurational time-dependent Hartree (MCTDH) method is employed, extending its application to few-fermion systems. Apart from numerical calculations we generalize our Ansatz for a correlated pair wave-function proposed in \cite{brouzos} for bosons to mixtures of fermions. From weak to strong coupling between the components the energies, the densities and the correlation properties of one-dimensional systems change vastly with an upper limit set by fermionization where for infinite repulsion all fermions can be mapped to identical ones. The numerical and analytical treatments are in good agreement with respect to the description of this crossover. We show that for equal populations each pair of different component atoms splits into two single peaks in the density while for partial imbalance additional peaks and plateaus arise for very strong interaction strengths. The case of a single impurity atom shows rich behaviour of the energy and density as we approach fermionization, and is directly connected to recent experiments \cite{selim,selim1,selim2}. 
\end{abstract}

\maketitle

\section{Introduction}

Ultracold Fermi gases have been a flourishing research field in the last decade after the early experiments on $^{40}$K \cite{K40} and $^6$Li \cite{Li6}. Impressive features like superfluidity according to the Bardeen-Cooper-Schrieffer (BCS) theory \cite{bcs} have been realized experimentally \cite{bcsexp}. Two-component ensembles prepared usually in two hyperfine states or with different atomic species are in the core of research for novel phases.  Ensembles with imbalanced populations \cite{imbalancedexp} can then lead to the coexistence of normal and superfluid phases \cite{imbalanced} or to the more complex phase with a spatially varying gap  as proposed by Fulde, Ferrell, Larkin and Ovchinnikov (FFLO) \cite{fflo}. Highly imbalanced Fermi mixtures constitute a very well controllable system to study polaron physics  and recently two experiments have achieved measurements of polaronic behaviour for attractive \cite{polaronexpatt} and repulsive impurity atoms \cite{polaronexprep} in good agreement with corresponding theoretical treatments \cite{polaron,polaronrep}. An open debate in the field is the  observation of itinerant ferromagnetism induced by interactions \cite{ferro}.

The regime of strong inter-atomic coupling is particularly interesting, since various spectacular properties of highly correlated phases arise. This holds in particular  for one-dimensional (1D) quantum gases which are strongly confined with respect to their transversal modes. These quasi-1D tubes are usually equipped with an external harmonic trap for the longitudinal direction. Apart from the well-established toolbox of cold atoms \cite{bloch} containing optical traps and magnetic Feshbach resonances to control the trapping geometry and the interaction strength respectively, there exist for quasi-1D setups also  the so-called confinement induced resonances \cite{olshanii} modifying the free scattering properties due to the presence of the confinement. One of the most peculiar features in one-dimension is the mapping between bosons and fermions for infinite repulsion which induces an effective Pauli exclusion principle to the bosonic species \cite{girardeau}. This fermionization of bosons, the so-called Tonks-Girardeau (TG) gas has been observed experimentally \cite{kinoshita} and more recently its highly excited counterpart for attractive interactions has been probed \cite{haller}. The fermionic physics in 1D may prove to be even richer in phenomenology as the first experiments in confinement-induced molecules \cite{molecule1d} and spin-imbalanced Fermi gases \cite{imbalanced1d} indicate. On theoretical side there have been already several studies on the corresponding polaronic physics \cite{McGuire,giraud,lamacraft,torma,guan}, TG and super-TG gases specifically their pairing properties and dynamics \cite{girardeau2,guanchen,girardeau3}. 

In this work, we investigate a few-fermion system composed of two species with the interspecies coupling strength being the control parameter which is tuned from weak to strong interactions. Methodologically we employ two approaches: (a) the numerically exact multi-configurational time-dependent Hartree (MCTDH) method \cite{meyer90} (see Appendix), which has been proven to be efficient for few-boson systems \cite{sascha}, applied here for first time on a fermionic few-body ensemble and (b) the generalization for mixtures of the explicit wave-function based on products of pair-functions which the authors introduced for identical bosons in Ref. \cite{brouzos}. The latter Ansatz has been proven to be efficient for the  description of  the few-body properties in the course of the crossover from weak interactions to fermionization, and the extension here to few-fermion two-component mixtured is also in a good agreement capturing the main behaviour of the observables. Our present study is in close connection to the experimental work \cite{selim}, which has achieved the deterministic loading of a specific number of fermions into a trap  with a very high fidelity \cite{selim1,selim2} and accurately controls the imbalance of the population. The state-of-the art of these experiments, allows for a study of specific configurations of populations of different hyperfine states like those which we aim at here: highly imbalanced with one impurity atom surrounded by several (1-4) majority atoms, a balanced system with equal populations of totally each species, and a partially imbalanced ensemble of eg. 3 majority to 2 minority atoms. For these cases we will study the energies and the densities as the repulsive interaction strength increases, showing thereby the mechanisms and features that characterize these few-body ensembles along the crossover to fermionization. The case of completely imbalanced population, is already measured from the experiment in a work to appear soon \cite{selim}.

This article is organized as follows: In Section II we present the setup and the explicitly correlated wave function which is proposed for its description. In the following Section III we analyze and compare the analytical and numerical results of the energies for all configurations of  few-fermion mixtures included in this work. Subsequently we report explicitly on properties of the one-body density for the impurity case (Sec. IV) and for the partially as well as completely balanced systems (Sec. V). In the last Section  VI we present our conclusions and an outlook. The Appendix contains a short description of the MCTDH method and remarks on its implementation for few-fermion systems. 

\section{Hamiltonian and Ansatz}

The system consists of a two component Fermi gas with the two species prepared e.g. in the two lowest hyperfine states ( for $^6$Li $|F=1/2, m_F= \pm 1/2>$ see ref. \cite{selim1,selim2}) confined in a pure 1D harmonic oscillator potential.  The standard experimental approach is to create quasi-1D tubes by using a very strong laser field for the transversal degrees of freedom compared to the longitudinal one \cite{kinoshita,haller}. The trap becomes then highly anisotropic with the characteristic transversal length scale  $a_{\perp} \equiv \sqrt {\frac{\hbar}{M \omega_{\perp}}}$ much smaller than the longitudinal one $a_{\parallel} \equiv \sqrt {\frac{\hbar}{M \omega_{\parallel}}}$ [$\omega_{\perp}$ ($\omega_{\parallel}$) is the transversal (longitudinal) harmonic confinement frequency]. The typical ratio of the confinement frequencies   $\omega_{\parallel}/ \omega_{\perp} \leq 1/10$ is adequate to characterize the system as 1D. Therefore the transverse degrees of freedom are energetically frozen occupying only the corresponding ground states.

Due to antisymmetry s-wave interactions between identical fermions (same hyperfine state) are not possible, while p-wave interactions are negligible. Therefore we remain with a single interaction parameter, namely the interaction strength that between atoms belonging to different components. Still this has to take into account the quasi-1D nature of the confining potential and the resulting confinement induced resonances \cite{olshanii}. The effective 1D interaction strength reads:
\begin{equation*}
g_{1D}= \frac{2\hbar^2 a_{3D}}{M a^2_{\perp}} \left(1-\frac{|\zeta(1/2)| a_{3D}}{\sqrt{2} a_{\perp}}\right)^{-1},
\end{equation*}
where $a_{3D}$ is the 3D s-wave scattering length. When $a_{3D}$ which can be tuned via magnetic Feshbach resonances approaches the transversal confinement length $a_{\perp}$, a confinement induced resonance is encountered. Therefore the inter-species interaction strength can be regarded as tunable over a very large range.

With this information at hand we may write the rescaled 1D harmonic oscillator Hamiltonian:
\begin{equation}
\label{ham}
H = - \frac{1}{2} \sum_{i=1}^N \frac{\partial^2}{\partial  x_i^2} + \frac{1}{2} \sum_{i=1}^N x_i^2 + g \sum_{i \leq N_M <j \leq N}\delta(x_i-x_j)
\end{equation}
such that lengths and energies are scaled by  $a_{\parallel}$ and  $\hbar\omega_{\parallel}$ respectively, and $g=g_{1D}/\hbar \omega_{\parallel} a_{\parallel}$. The total number of atoms is $N$ and $N_M$ is number of atoms belonging to the hyperfine state which has the larger population in the trap. We remain therefore with a single parameter $g$, the rescaled coupling strength between the $N_M$ majority fermions and the $N_m=N-N_M$ minority fermions. We will examine in the next sections the following cases
\begin{itemize}
 \item a single impurity for the completely imbalanced configurations 2:1, 3:1, 4:1
 \item a balanced population 2:2
 \item a partially imbalanced population 3:2
\end{itemize}
the notation being $N_M$:$N_m$. We restrict ourselves to these few-body scenarios which are relevant to the experiment \cite{selim1,selim2} (the first case is already realized experimentally \cite{selim}) and for which we can obtain reliable accurate numerical results with the MCTDH method (see Appendix). In most cases these are sufficient to reveal the mechanisms present also for larger $N$. 
  
The special balanced case of two particles (1:1) allows for an exact solution not only in 1D but also in higher dimensions analyzed in the seminal paper of Busch et. al. \cite{busch}.  The energy curves obtained there solving a transcendental equation, were also measured experimentally in ref. \cite{selim2} with a very good accuracy via tunneling rates after opening the harmonic trap from one side. Extending this study to higher particle numbers from the theoretical side is the basic scope of this work. In this context, the authors have proposed in ref. \cite{brouzos} an explicitly correlated pair wave-function (CPWF) for identical bosons which is based on a construction principle inspired by the two-body solution \cite{busch}. This CPWF will be extended and generalized here to cover the case of two-component Fermi mixture.

The Hamiltonian [Eq. (\ref{ham})] containing the harmonic trap potential can be separated into a center-of-mass and relative motion part. The center of mass $R=\sum_i^N x_i/N$ obeys exclusively in its harmonic oscillator Hamiltonian, with the ground state $\Psi_{CM}=(N/\pi)^{1/4}e^{-NR^2/2}$. The equation of the relative motion in the general case is much more complex, contains kinetic couplings of the relative degrees of freedom, includes the interaction terms and does not allow for an analytical solution. The CPWF which we proposed \cite{brouzos} in the case of bosons for the relative motion of $N$ particles consists of products of parabolic cylinder functions (PCF) $D_{\mu}$ \cite{abramowitz}, which is the key function occurring in the solution of the two-body problem, of the relative coordinate $r_{ij}=x_i-x_j$ of each pair. The substantial difference in the case of fermions is that the pairs of identical particles (atoms belonging to the same hyperfine states) should be put on the levels of the Fermi ladder as we will see in the following. 

The generalization of the CPWF we propose here for the relative motion reads then: 
\begin{equation}
\label{psi}
 \Psi_{\mathrm{cp}} = \prod_{i \leq N_M <j }^{P}D_{\mu}(\beta |r_{ij}|) \prod_{i <j \leq N_M}^{P_M} D_{1}(\beta r_{ij}) \prod_{N_M< i <j}^{P_m}D_{1}(\beta r_{ij})
\end{equation}
(up to a normalization constant which we neglect here), where $P$ is the number of distinct pairs of atoms belonging to different hyperfine states, $P_M$ and $P_m$ are the number of pairs of majority and  minority atoms respectively. The pinning of identical fermions belonging to the same component on the corresponding Fermi-ladder is done by choosing $\mu=1$ as a parameter for the PCFs (see the two last terms of Eq. \ref{psi}) which corresponds to the energetically lowest orbital of the fermionic relative motion: $D_{1}(x) =\sqrt{2} x e^{-x^2/4}$. Therefore the part of the total wave function which refers to identical fermions is the correct non-interacting ground state resulting from the Slatter Determinant: $\Psi_{id}= \prod_{i} e^{-x_{i}^2/2} \prod_{i < j }  (x_i-x_j)$ with all $x_i$'s and $x_j$'s being exclusively coordinates belonging to the same species of fermions here. For the interacting pairs in the wave function the parameter $\mu$ of the PCF (see first term of Eq. \ref{psi}) which depends on the interaction strength $g$ should be determined. This can be done exploiting the condition $2 \beta D'_{ij}(0)= g D_{ij}(0)$, where  $D'_{ij}=\frac{\partial D_{ij}}{\partial ( \beta r_{ij})}$ and $D_{ij} = D_{\mu}(\beta r_{ij})$, which is imposed for each pair of atoms belonging to different components in the ensemble at $r_{ij}=0$ (as we have shown in Ref. \cite{brouzos}) due to the discontinuity that is imposed by  the delta-like interaction potential on first derivative. The resulting transcendental equation 
\begin{equation}
\label{bc}
\frac{g}{\beta}=-\frac{2^\frac{3}{2}\Gamma\left(\frac{1-\mu}{2}\right)}{\Gamma\left(\frac{-\mu}{2}\right)}
\end{equation}
is solved for $\mu$, selecting the solution in the interval $\mu \in [0,1]$ which corresponds to the ground state. 

Another key feature of the CPWF is that it reproduces the exact analytically known solutions in the limits $g=0$ and $g \to \infty$, for any $N$ by choosing $\beta=\sqrt{\frac{2}{N}}$.  In these limits, $\mu$ equals $0$ and $1$, respectively, and in both cases (as well as for every integer $\mu$) $D_{\mu}(x)=e^{-\frac{x^2}{4}}\mathrm{He}_{\mu}(x)$ where $\mathrm{He}_{\mu}(x)$ are the modified Hermite polynomials. Therefore the total wave function takes the following form  
\begin{equation}
\label{nonint}
\displaystyle\Psi_{g=0}= e^{-\sum_{i=1}^{N} x_i^2/2}\prod_{i <j \leq N_M}^{P_M} r_{ij} \prod_{ N_M< i <j}^{P_m} r_{ij}
\end{equation}
 for the non-interacting and 
\begin{equation}
\label{tg}
\displaystyle\Psi_{g \to \infty}=  e^{-\sum_{i=1}^{N} x_i^2/2}\prod_{i \leq N_M <j }^{P}|r_{ij}|\prod_{i <j \leq N_M}^{P_M} r_{ij} \prod_{N_M<i<j}^{P_m} r_{ij}
\end{equation}
for the infinitely strong repulsion limit \cite{girardeau,girardeau2,girardeau3} which are the exact solutions in both cases. The CPWF is also exact for arbitrary $g$ for the case $N=2$ (with $\beta=1$) \cite{busch} and indeed our general Ansatz is inspired by this two atom case \cite{busch,brouzos}. 

Using the CPWF as trial function one can treat $\beta$ as a variational parameter for given values of $g$ and $N$. The form of the pair-correlated function is reminiscent of the Bijl-Jastrow form of products, the  main difference being here that we use the exact solution of the two atom case instead of a trigonometric function which is the usual approach e.g. in quantum Monte Carlo studies. Generalizations of this function can also be extended to other cases like bosonic mixtures or even fermion-boson mixtures or to many component mixtures. One should take $\mu=1$ for identical fermions and a   $\mu \in [0,1]$ for interacting pairs according to Eq. \ref{bc}. Yet one has to keep in mind that in case of more than two species and for the situation that only a subset of the coupling strengths  are infinite the above Ansatz may not be exact. Consequently the cases of a single species bosons and two species fermions are the most natural ones for which this Ansatz should have its best performance. Another possible generalization is to separate the PCF into a product of Hypergeometric and exponential functions $D_{\mu}(x)= 2^{\frac{\mu}{2}}e^{-\frac{x^2}{4}}U(-\frac{\mu}{2},\frac{1}{2},\frac{x^2}{2})$,  and choose two independent variational parameters one in the exponent and one in the confluent hypergeometric function of second kind $U(a,b,x)$ \cite{abramowitz}. We will restrict ourselves here to the approach we addressed above for two-component fermions (we will only treat a single case (2:1) variationally with a single parameter $\beta$) in order to show the efficiency of CPWF as it stands in Eq. \ref{psi} to describe the crossover from weak to strong interactions. 

A first illustration of the behaviour of the CPWF is given for the case 2:1 in Fig. 1 (b) for weak and (c) for strong interaction strength, where contour plots of the probability density are shown. On the manifolds $\mathcal{M}_{ij}=\{(x_1,...x_i,...,x_j,...,x_N) \in \mathbb{R}^N |x_i=x_j \}$ where the identical atoms meet (here the plane $x_1=x_2$) the probability density possesses already for weak interactions [Fig. 1 (b)] a low probability, while stronger interactions between the non-identical atoms  [see Fig. 1 (c)] reduce the values of the density also on the other two planes ($x_1=x_3$, $x_2=x_3$). This is a main intuitive picture that the CPWF captures, treating each fermionic pair --being interacting or not -- correctly at its contact point, where the wave function should tend to vanish as we increase the coupling strength from weak to strong interactions. In the next section the accuracy of this approach with respect to the energy is tested by comparison with corresponding numerical results for all cases examined in this work. The specific behaviour of the densities will be addressed in the  sections thereafter. 

\begin{figure*}
\includegraphics[width=12 cm,height=12 cm]{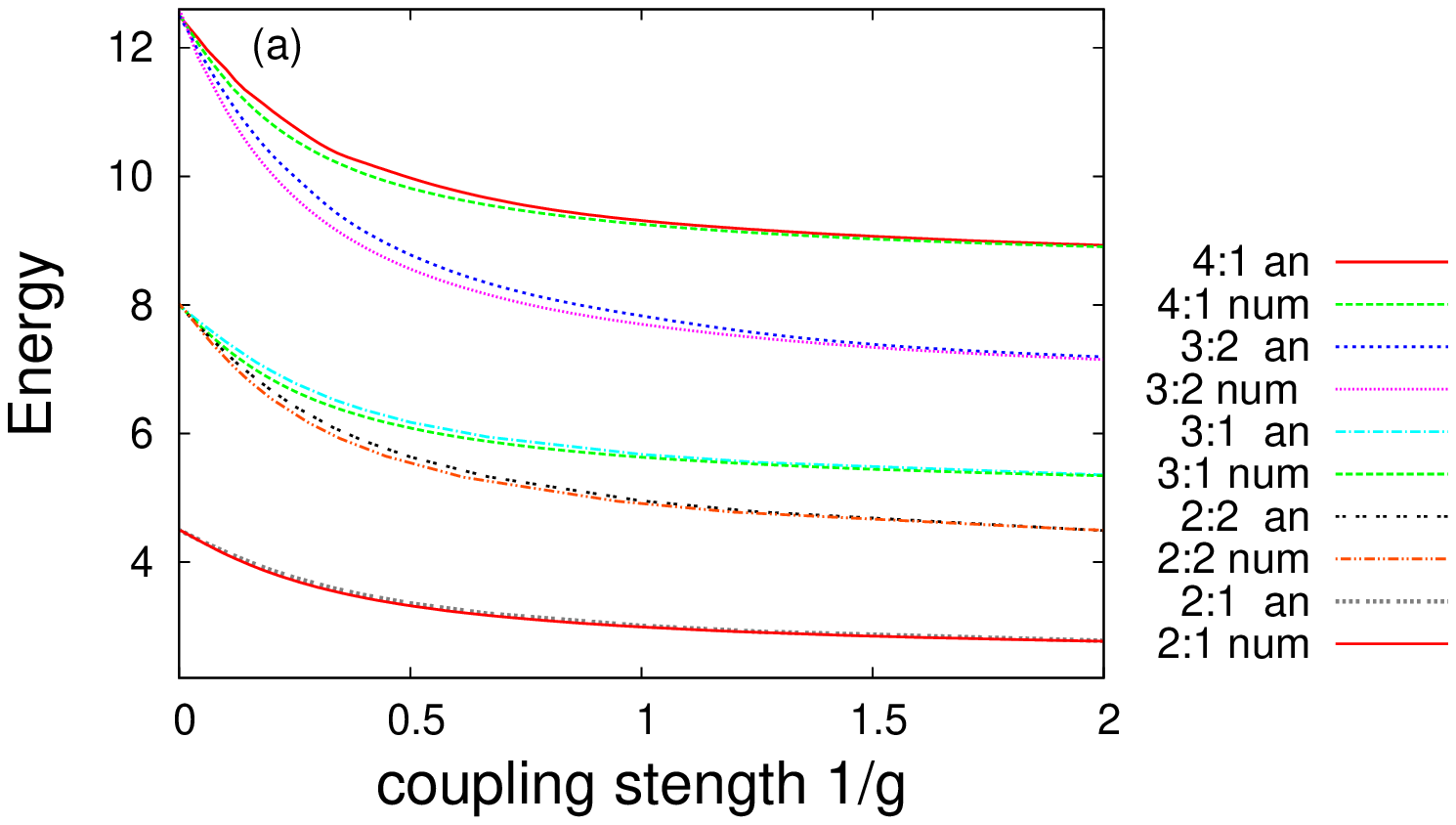}\\
\includegraphics[width=8.0 cm,height=8.0 cm]{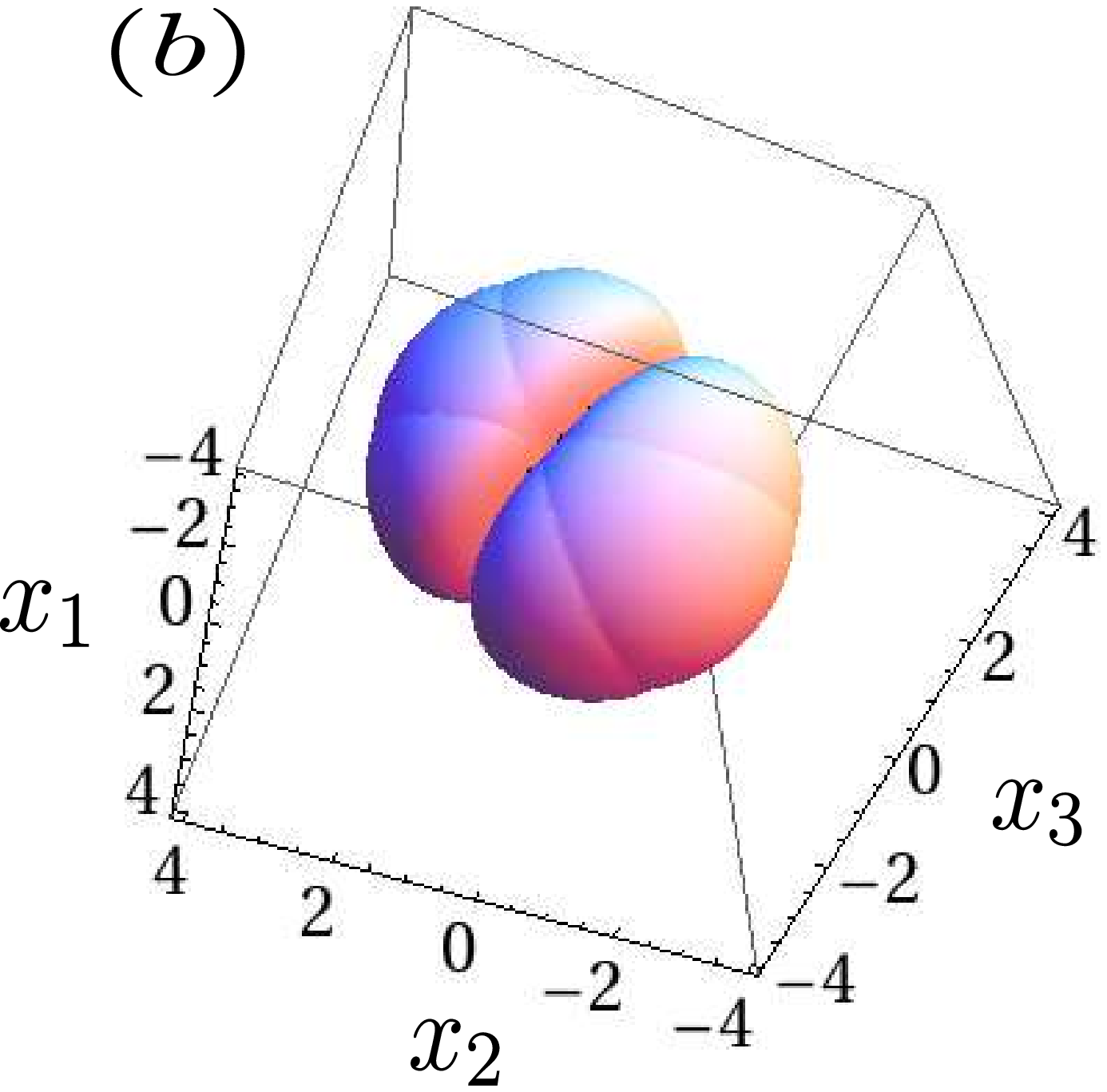}
\includegraphics[width=8.0 cm,height=8.0 cm]{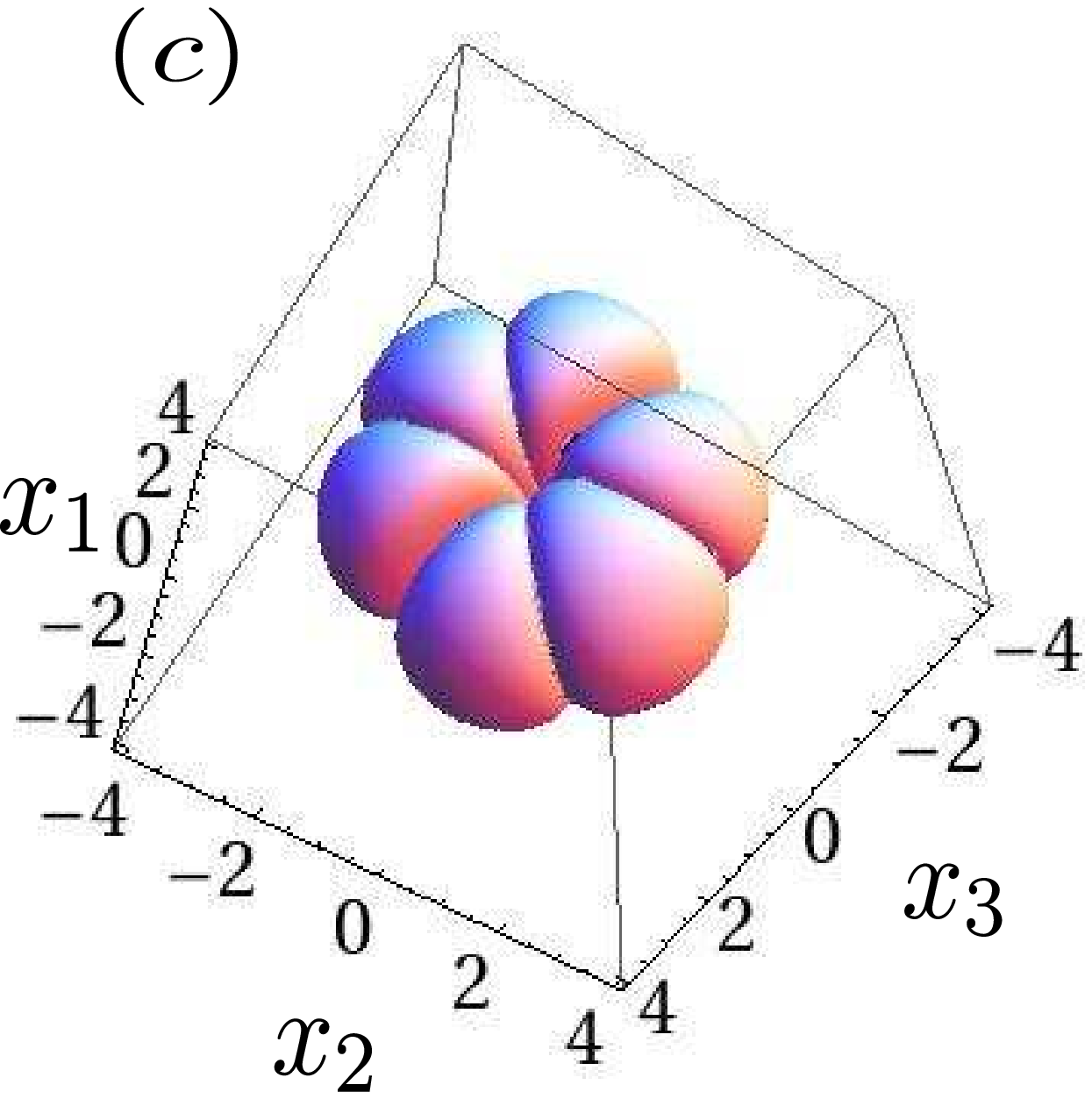}
\caption{(a) Energy as a function of the inverse interaction strength $1/g$ for all cases examined in this work (impurity 2:1, 3:1, 4:1, balanced 2:2, imbalanced 3:2). For each case the lower line corresponds to the numerical results employing MCTDH and the  line slightly above to the energy computed from the Ansatz CPWF Eq. \ref{psi}.  Contour plots of the density distribution $|\Psi(x_1,x_2,x_3)|^2=0.01$  for the configuration 2:1 using the CPWF for a weak (b) $g=0.5$ and a strong (c) $g=5.0$ interaction strength.}
\label{fig1}
\end{figure*}

\section{energies and accuracy of the Ansatz}

The energy of the interacting ensemble is a first observable that can be measured experimentally for example by RF-spectroscopy (see e.g. Refs. \cite{polaronexpatt,polaronexprep,selim}) or by determining tunneling rates \cite{selim2}. We compare here the numerical MCTDH results to the calculations using the CPWF and consequently obtain information on the validity and accuracy of the CPWF.

As discussed in the Appendix MCTDH always involves a cut-off for the number of single-particle functions which is employed, which in turn imposes also restrictions to the number of particles that can be treated with sufficient accuracy. Most  numerical methods are based on a truncated Hilbert space, and therefore result in inaccuracies especially when the system is highly correlated and a few (even optimal) orbitals are not enough to describe it. For the 1D case we examine here, there is an efficient workaround to overcome this problem, which is related to the fact that the energy should converge for $g \to \infty$ to a certain value equal to the one of the corresponding system composed of identical fermions. In Ref. \cite{dieter} a detailed study of how to overcome the truncated Hilbert space problem was performed for a bosonic system in the harmonic trap. The main trick is that for a certain set of values for the numerical parameters (number of basis functions and grid points) one has to rescale $g$ such that the energy becomes equal to the fermionic one only at $g \to \infty$. Therefore if $g_0$ is the finite value where the energy of the numerical calculation reaches fermionization, the corrected rescaled $g$ should be $g=\frac{g_n}{1-g_n/g_0}$ where $g_n$ is the coupling strength used in the numerical calculation. This procedure is available and of importance also to our case of two-component fermion system, and we use it for extracting the numerically exact energy curves shown in Fig. 1 (a) as a function of the inverse interaction strength (analogous to the effective 1D scattering length, see ref. \cite{olshanii}).

In Fig. 1 we also show the energies obtained from calculating the expectation value of the Hamiltonian operator using the CPWF [Eq. (\ref{psi})]. The accuracy of the Ansatz is very good especially for weak and very strong interactions, yet there are deviations from the numerical results in particular for intermediate coupling strengths, especially for larger particle numbers (see e.g. the curves for the configuration 2:1 in comparison with 4:1). It is to be expected that the CPWF works good close to the two exactly known limiting cases since it reproduces them [see Eqs. (\ref{nonint}) and (\ref{tg})]. The deviation for larger numbers of atoms is due to the two-body nature of the function which does not take into account higher correlations. Nevertheless a relative error of 2-3$\%$ can be achieved also by taking variational trial functions with a single variational parameter. 

Concerning the physical behaviou, the energy increases as the coupling strength gets larger (Fig. 1), but not with the same slope for all configurations. The range of the energies is bounded by the non-interacting values (for the configurations 2:1, 2:2, 3:1, 3:2 and 4:1 it is 2.5, 4, 5, 6.5 and 8.5 respectively) and the value at fermionization which is $E_{g \to \infty}=N^2/2$. This means that e.g. the partially imbalanced case 3:2 or the equal population one 2:2 increase stronger than the corresponding impurity cases 4:1 and 3:1 since  more interacting pairs of atoms are included in the ensemble. On the other hand if we consider only the impurity cases, then the gain of energy per majority atom (compared to the corresponding non-interacting state)  at a finite value of the interaction strength is larger for a smaller number of atoms. This means that by adding majority fermionic atoms the energy gain per atom decreases since the majority atoms lie higher and higher in energy, such that they almost do not feel the impurity fermion at the bottom.  Yet when the interaction becomes very large the impurity affects all atoms and at fermionization the energy gain is for all cases equal to $E_{g \to \infty}-E_{g=0}=N^2/2-N_M^2/2-1/2=N_M$. Further features of the behaviour of the physics of the impurity in the finite Fermi-sea confined in a harmonic trap, a case also related to the experiment \cite{selim} will be illustrated in the next section. A relevant question in this context is the connection between many-body polaronic physics in the continuum (see eg. Refs. \cite{McGuire,giraud,polaron,polaronexpatt,polaronexprep}) and the few-body behaviour of an impurity in the harmonic trap.  

\section{An impurity atom in a sea of majority fermions}

\begin{figure*}
\includegraphics[width=6.0 cm,height=6.0 cm]{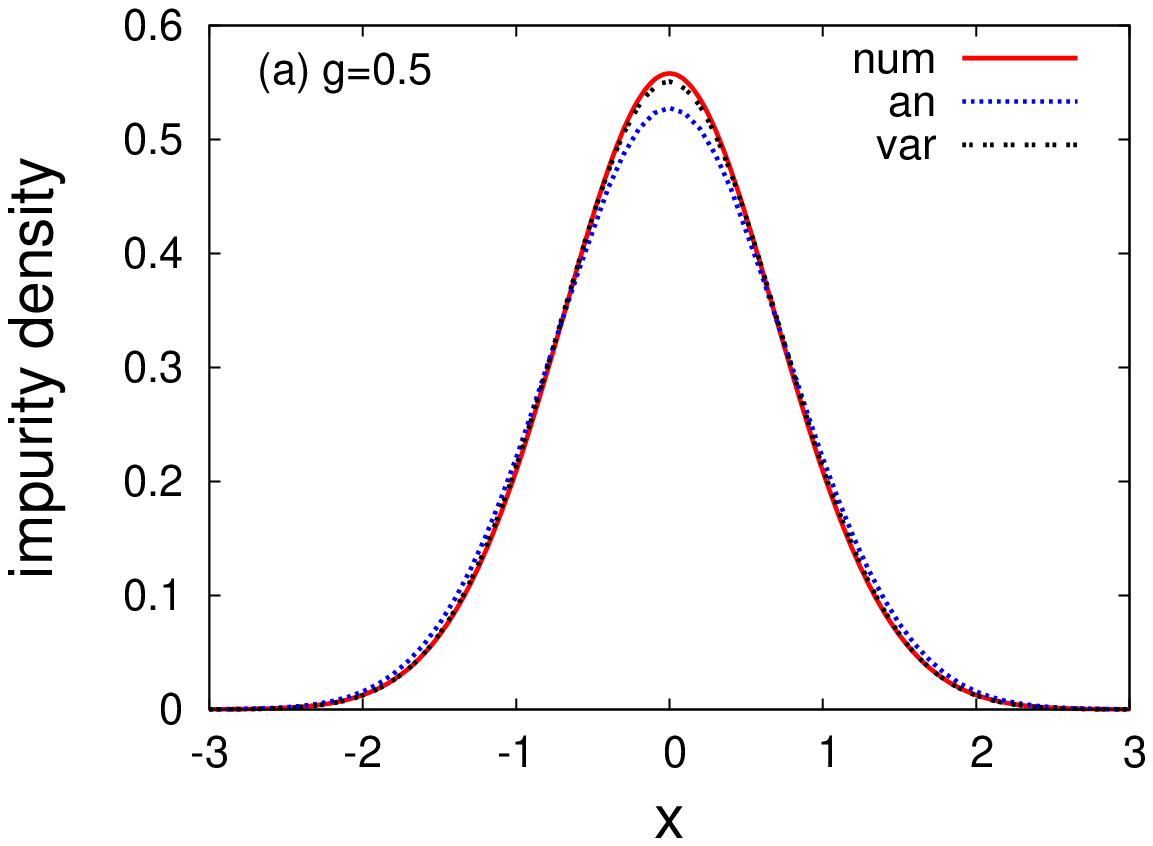}
\includegraphics[width=6.0 cm,height=6.0 cm]{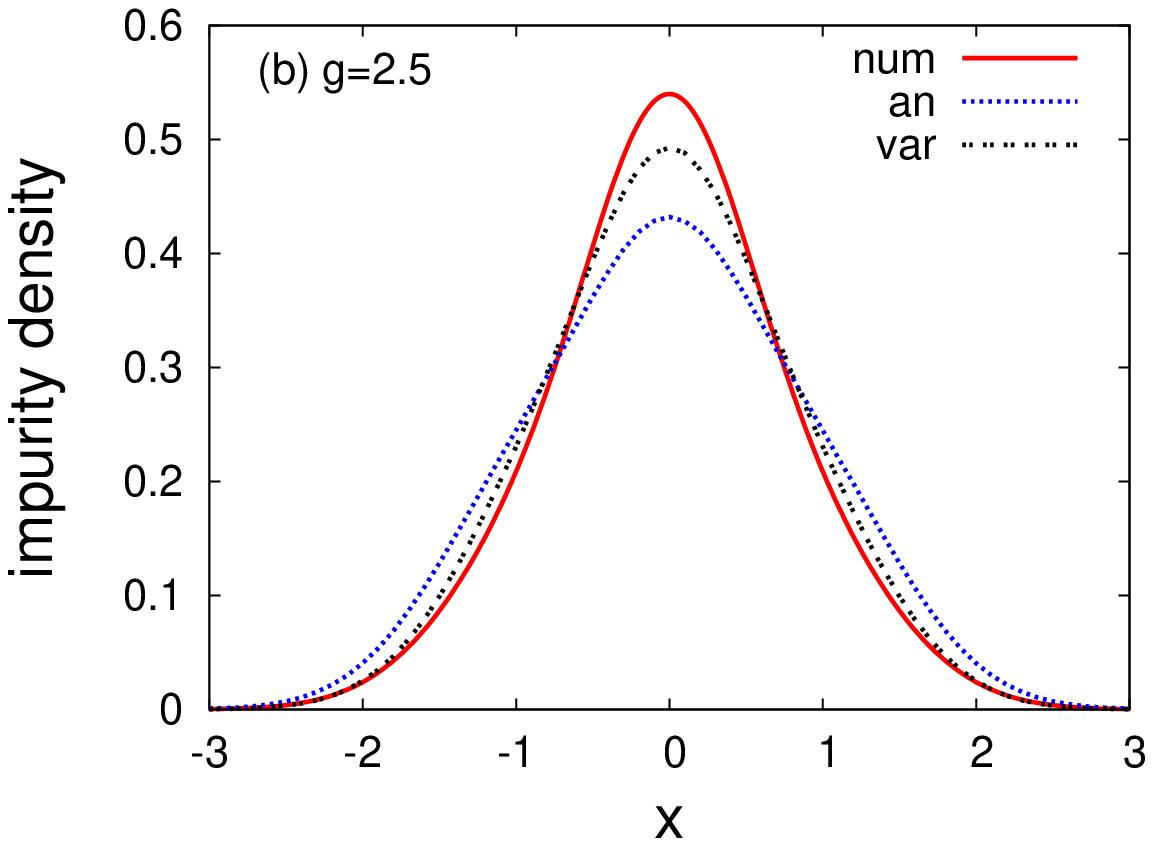}
\includegraphics[width=6.0 cm,height=6.0 cm]{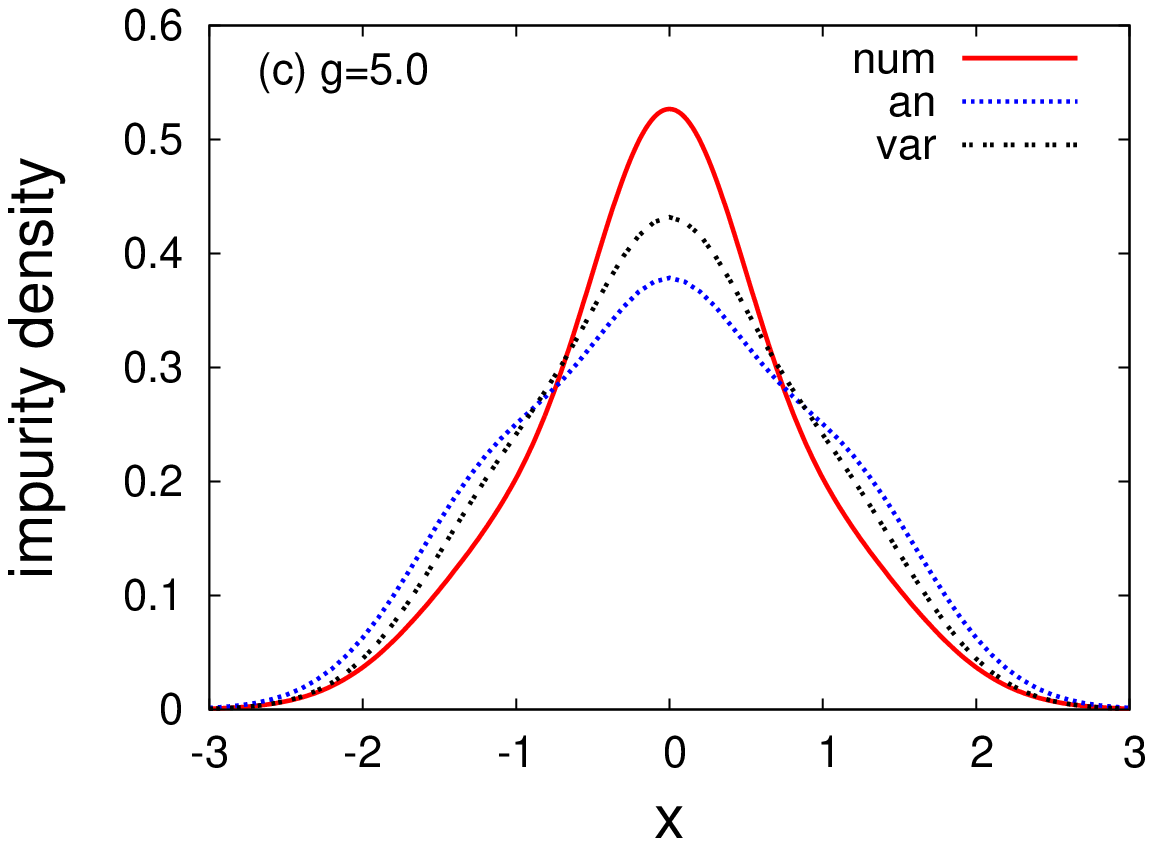}
\includegraphics[width=6.0 cm,height=6.0 cm]{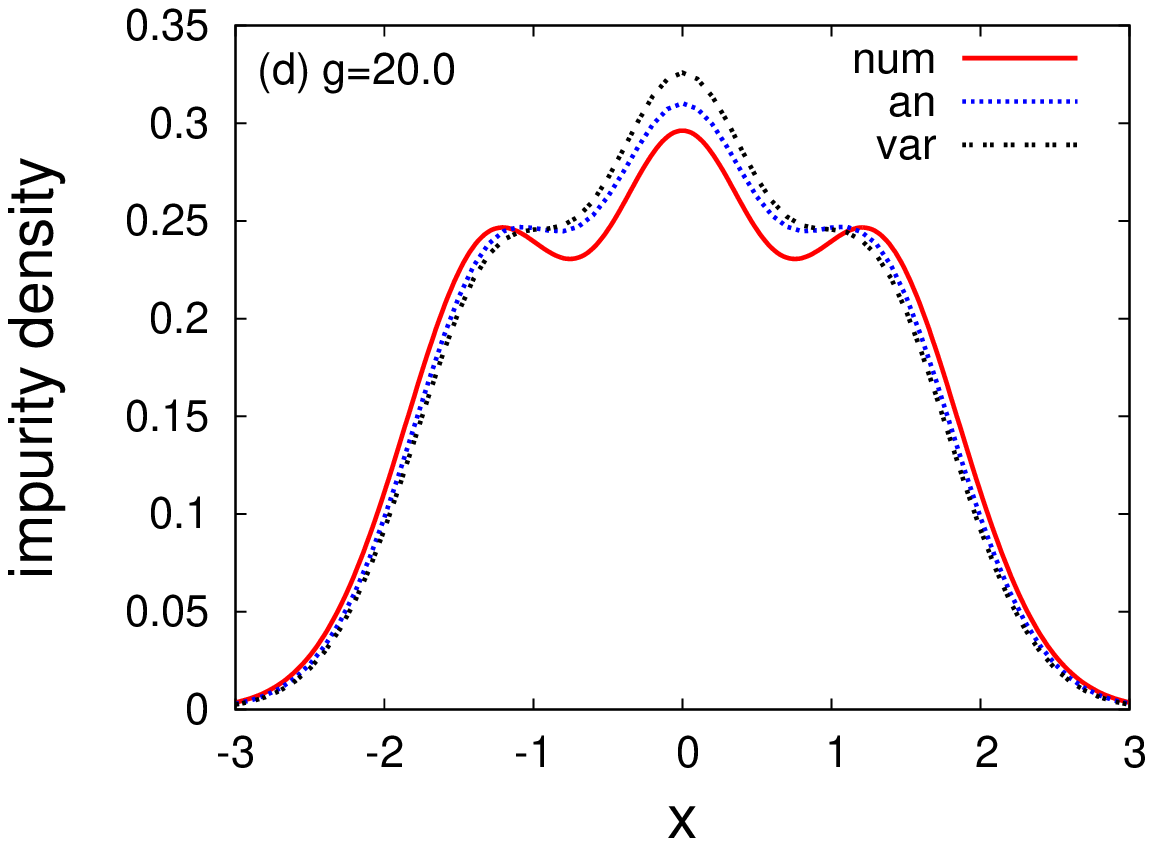}
\includegraphics[width=6.0 cm,height=6.0 cm]{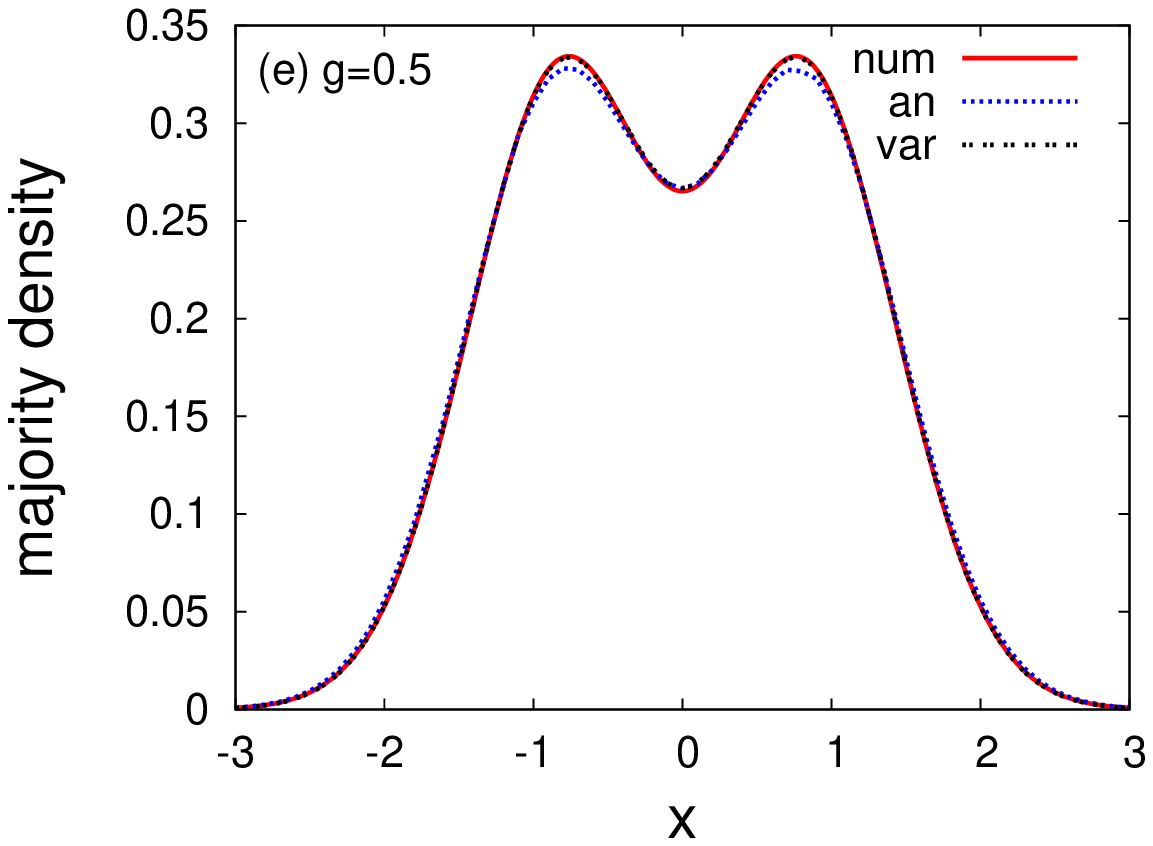}
\includegraphics[width=6.0 cm,height=6.0 cm]{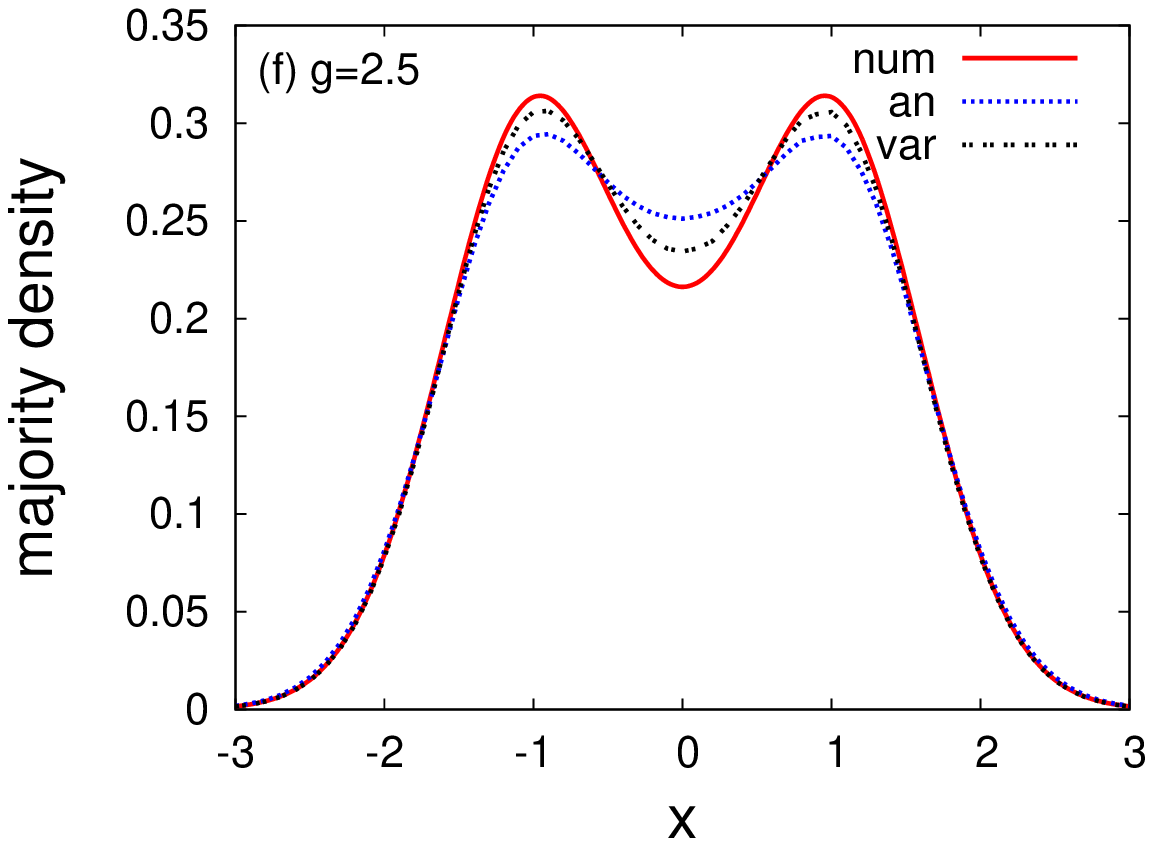}
\includegraphics[width=6.0 cm,height=6.0 cm]{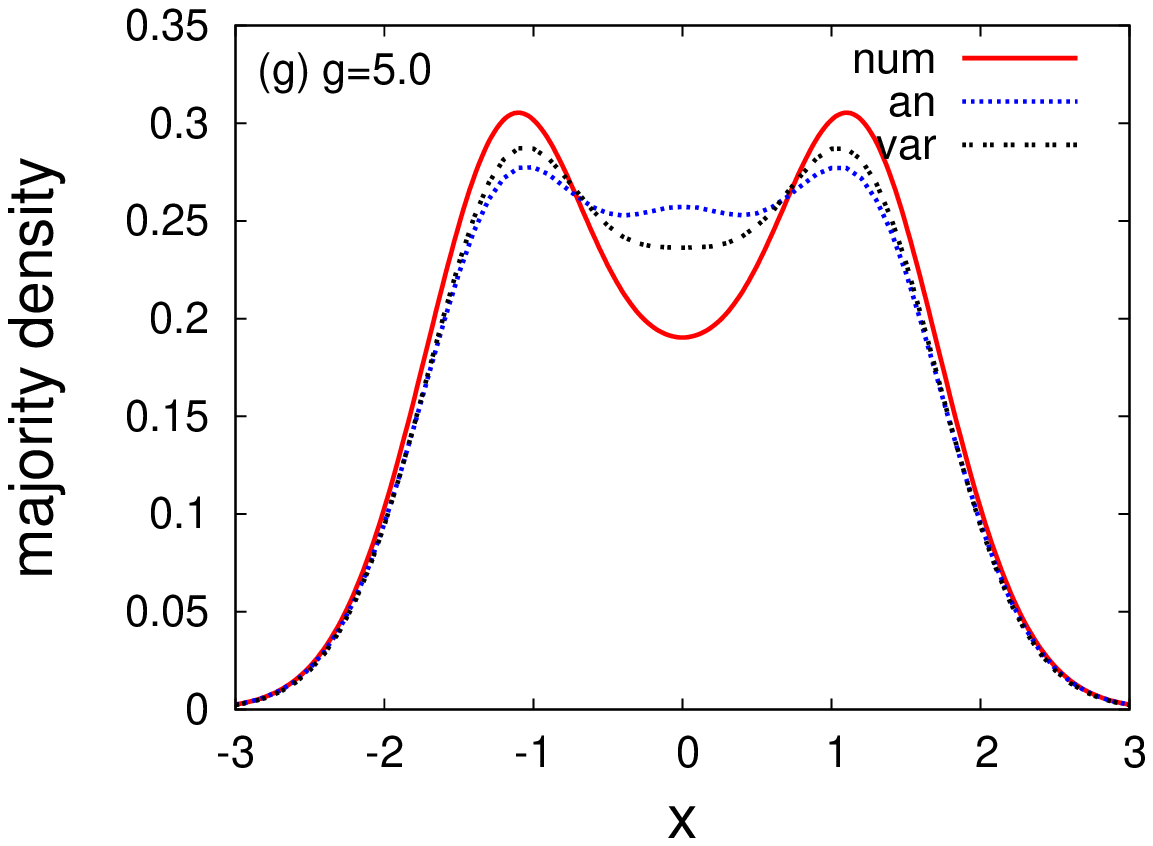}
\includegraphics[width=6.0 cm,height=6.0 cm]{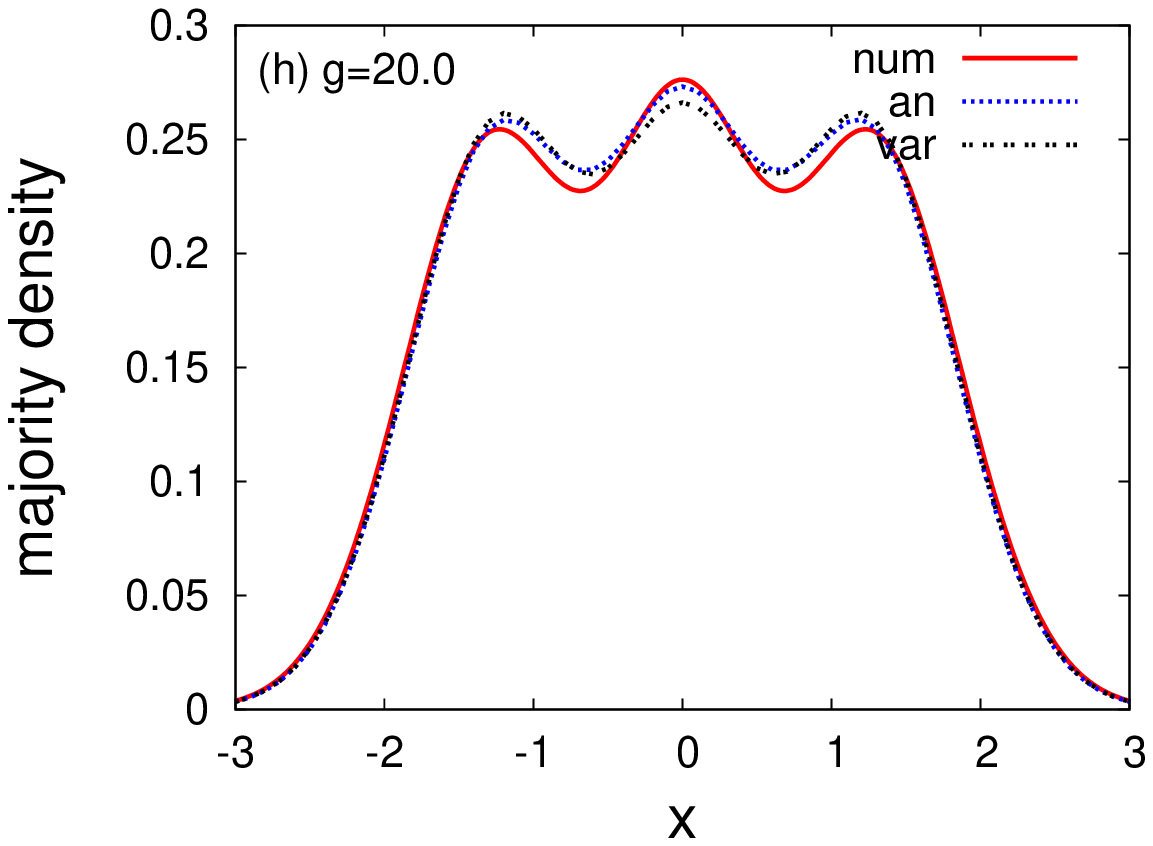}
\caption{One body densities of (a-d) the impurity  and  (e-h) majority fermions for the 2:1 case with varying interaction strength $g$ using the numerical (MCTDH) approach, and the analytical (CPWF) and variational Ansatz.}
\label{fig2}
\end{figure*}

In this section we focus on the impurity physics in 1D and analyze the corresponding probability densities. Major progress in recent experiments \cite{singleatomprobe} have extended towards probing the density distribution on a single site level. As demonstrated in Figs. 2 and 3 for the configurations 2:1 and 3:1 respectively, the probability density shows a vast change for both the majority particles and the single minority one. In the same Figures we show also the one-body density obtained by the CPWF (Eq. \ref{psi}). 

We focus first on Fig. 2 (a-d) which shows the evolution of the one-body density of the impurity from weak ($g=0.5, 2.5$) to stronger interactions ($g=5.0,20.0$). For weak interactions [$g=0.5$ Fig. 2 (a)] the impurity density is very close to the ground state orbital of the harmonic trap (Gaussian profile). As the interaction increases [$g=2.5$ Fig. 2 (b)] the density becomes broader and the peak lowers, which is a typical effect of repulsive interaction also in the case of bosons \cite{brouzos,sascha}. Still one may expect that in the case of fermions this single impurity atom would be even more confined and pinned in the middle of the trap due to the repulsive interaction with the surrounding majority atoms. However the opposite effect is observed: the density is stretched to cover a larger space in the trap and becomes delocalised. For even higher interaction strengths [$g=5.0$ Fig. 2 (c)] it even tends to acquire side lobes. More spectacularly for very strong interactions, the density of the impurity acquires three distinct maxima, an effect which is very evident in few-body fermionization profiles also for bosons \cite{girardeau1,brouzos,sascha}. The number of maxima is equal to the total number of atoms in the trap and according to Girardeau's theorem \cite{girardeau} the local properties like densities should be  exactly the same for a fermionized ensemble and for the corresponding ensemble of $N$ identical fermions. This vast change of the density of the impurity atom as we follow the crossover from weak interactions to fermionization also for the case of the 3:1 configuration is shown in Figs. 3 (a-d). 

The impurity is strongly affected by the repulsion with the majority atoms, which in turn are also strongly influenced by the single minority atom. Their densities for the case 2:1 depicted on Fig. 2 (e-h) possess at weak interactions  [$g=0.5$ Fig. 2 (e)] two maxima (corresponding to a two identical-fermion ground-state profile) which are pushed to outwards as the repulsion increases  [$g=2.5$ Fig. 2 (f), remember that the corresponding  impurity density becomes more delocalized in Fig. 2 (a),(b)] and start to acquire a plateau and finally an additional peak in the middle for very strong coupling  [$g=5.0, 20.0$ Fig. 2 (g),(h)]. A similar behaviour is observed also for the case 3:1 where the middle peak for weak interactions is slightly broadened pushing the outer maxima outwards [$g=0.5,2.5,5.0$ Fig. 3 (e),(f),(g)] and finally is turned to two peaks reaching the fermionization profile for strong coupling  [$g=20.0$ Fig. 3 (h)]. The impurity fermion affects therefore the density profile of the majority atoms equally strong, leading to the extreme infinitely repulsive limit where it becomes effectively indistinguishable from the majority fermions (strictly only for the local properties, of course). 

An issue that is worth noting here is that the very strong change in the density profile (where more distinct peaks appear) is happening rather abruptly for a very high interaction strength. The latter distinguishes the case of few fermions from that of few bosons which  acquire substantial peaks  already for intermediate interactions (see e.g. \cite{brouzos}). In other words, this is a characteristic behaviour of two-composite few fermion systems, which becomes interesting in the very strong correlation regime, being the important regime of 1D physics. This regime is difficult to access both for the experiment (one has to approach very closely to the $g \to \infty$ resonance, and simultaneously avoid losses to the molecular side \cite{selim}) and for numerical calculations since a lot of basis functions have to be included, rendering the treatment of a resulting huge Hilbert space very difficult. The application and convergence of MCTDH for such strong interactions is a very subtle issue, and has been tested extensively in this work, leading to the reliable results for the cases presented, which we also compare with the corresponding CPWF (which naturally provides the correct profile for infinite repulsion, thus not running into the afforementioned difficulties). We note that here the values of g are the actual ones used in the calculation and are not rescaled like in the case of the energy. 

The accuracy of CPWF to describe the evolution of the densities of the impurity or the majority particles is very good close to the two, the non-interacting and fermionization, limits as expected and describes the crossover qualitatively very well. Yet at intermediate interactions, the CPWF densities tend to be broader and acquire minor additional peaks compared to the numerical results  [see Figs. 2, 3 (b),(c),(f),(g)]. This is in accordance with behaviour of the corresponding energies, i.e., the fact that the analytical CPWF energies lie above the numerical ones in the intermediate interaction regime [see Fig. 1(a)]. For the 2:1 case  we present also the densities obtained by a variational treatment of the parameter $\beta$ (minimization of the energy using the CPWF  $\Psi_{cp}$ as a trial function). We observe in Fig. 2 that the variational treatment of the CPWF represents a significant improvement over the fixed $\beta$ CPWF case: the corresponding densities get closer to the numerical ones. Further improvements of the Ansatz to capture the exact quantitative behaviour may include more variational parameters or  a different construction in terms of hypergeometric functions discussed in section II. Yet we may emphasize that the correct qualitative behaviour is captured by the explicit function $\Psi_{cp}$. A similar good agreement exists for the next two cases of partial and complete balance of populations that we present in the next section.

\begin{figure*}
\includegraphics[width=6.0 cm,height=6.0 cm]{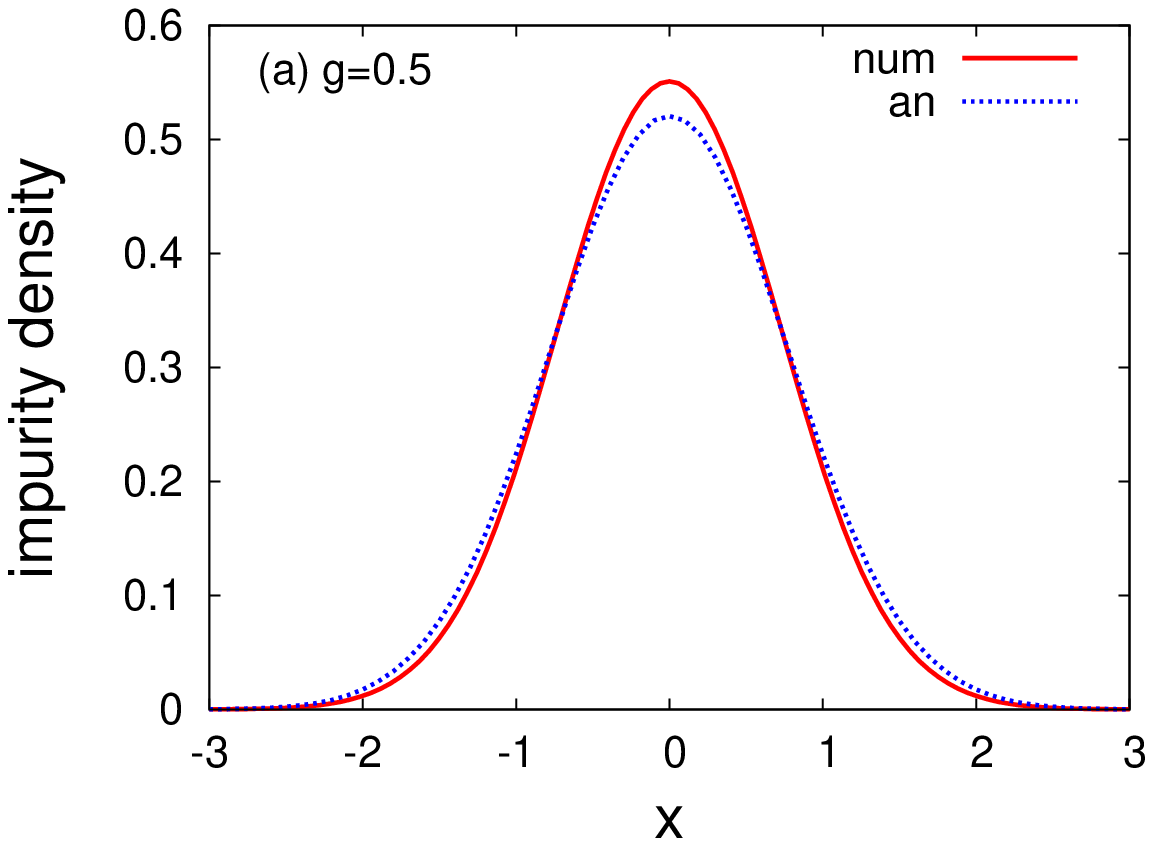}
\includegraphics[width=6.0 cm,height=6.0 cm]{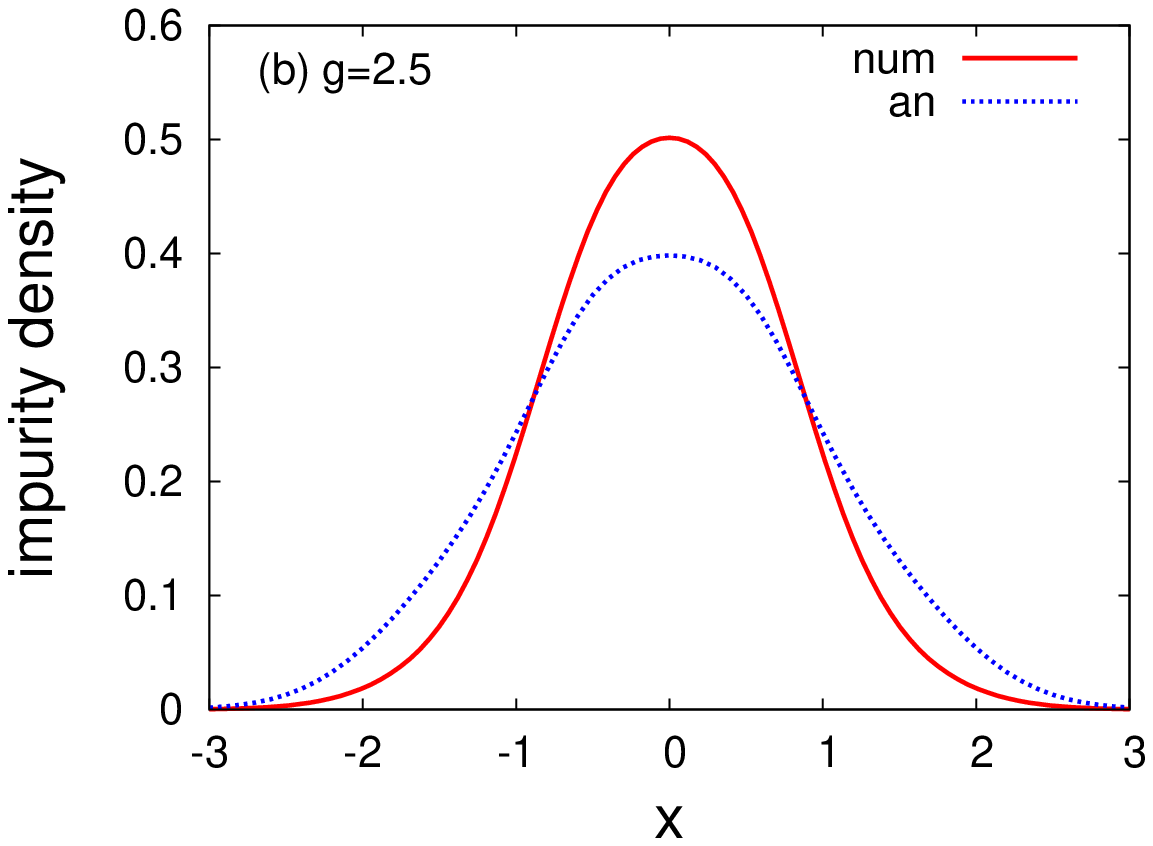}
\includegraphics[width=6.0 cm,height=6.0 cm]{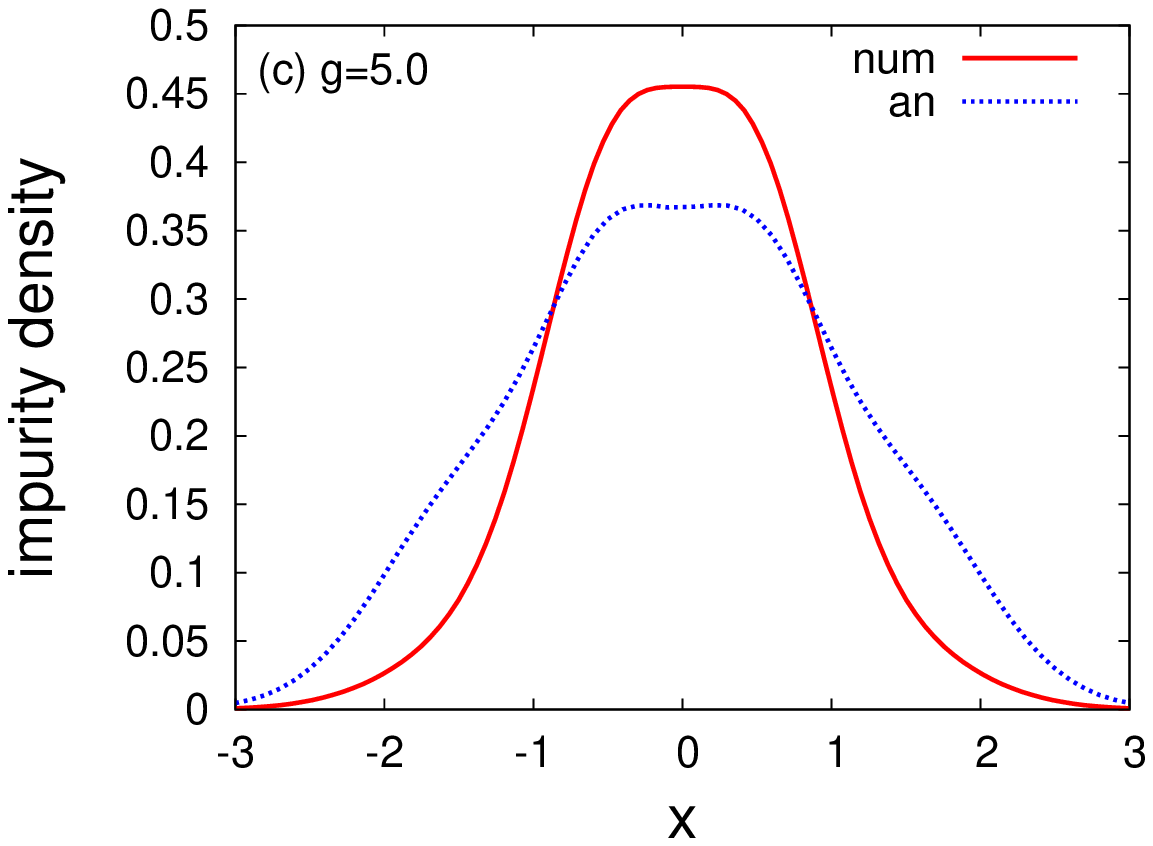}
\includegraphics[width=6.0 cm,height=6.0 cm]{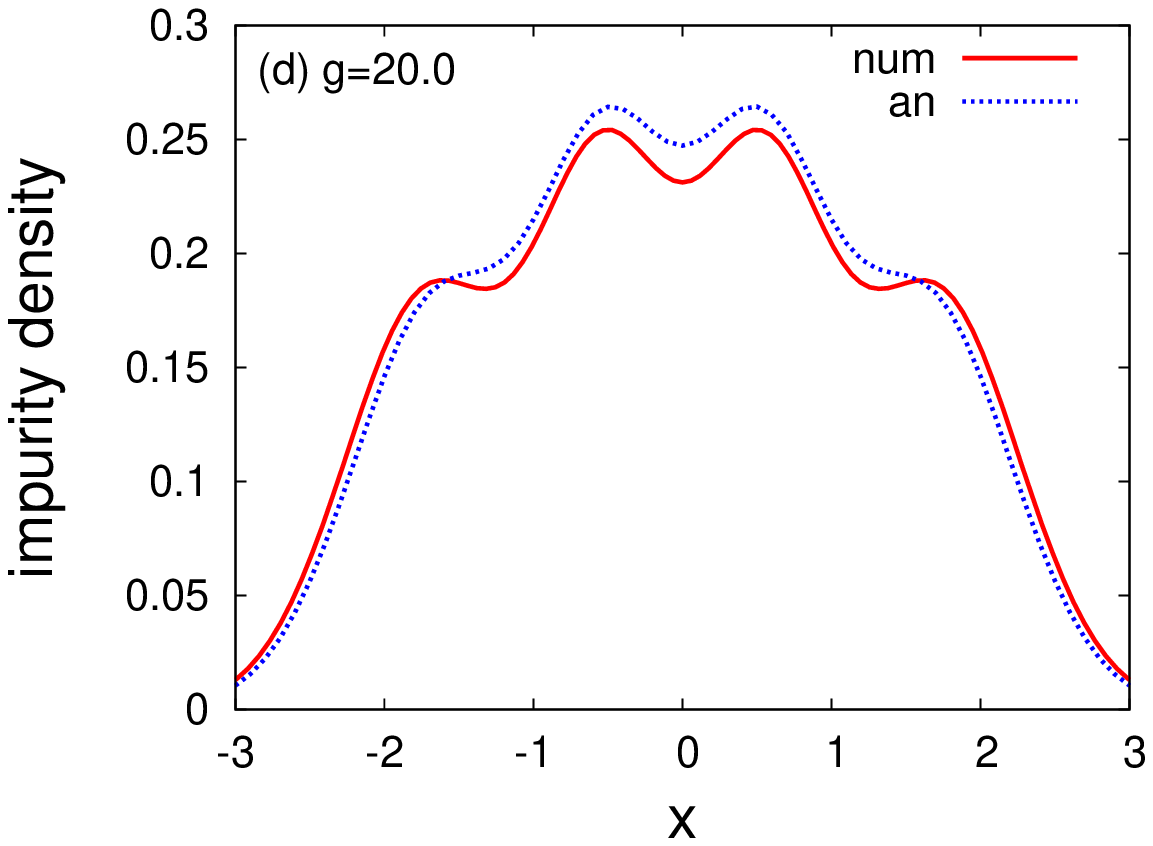}
\includegraphics[width=6.0 cm,height=6.0 cm]{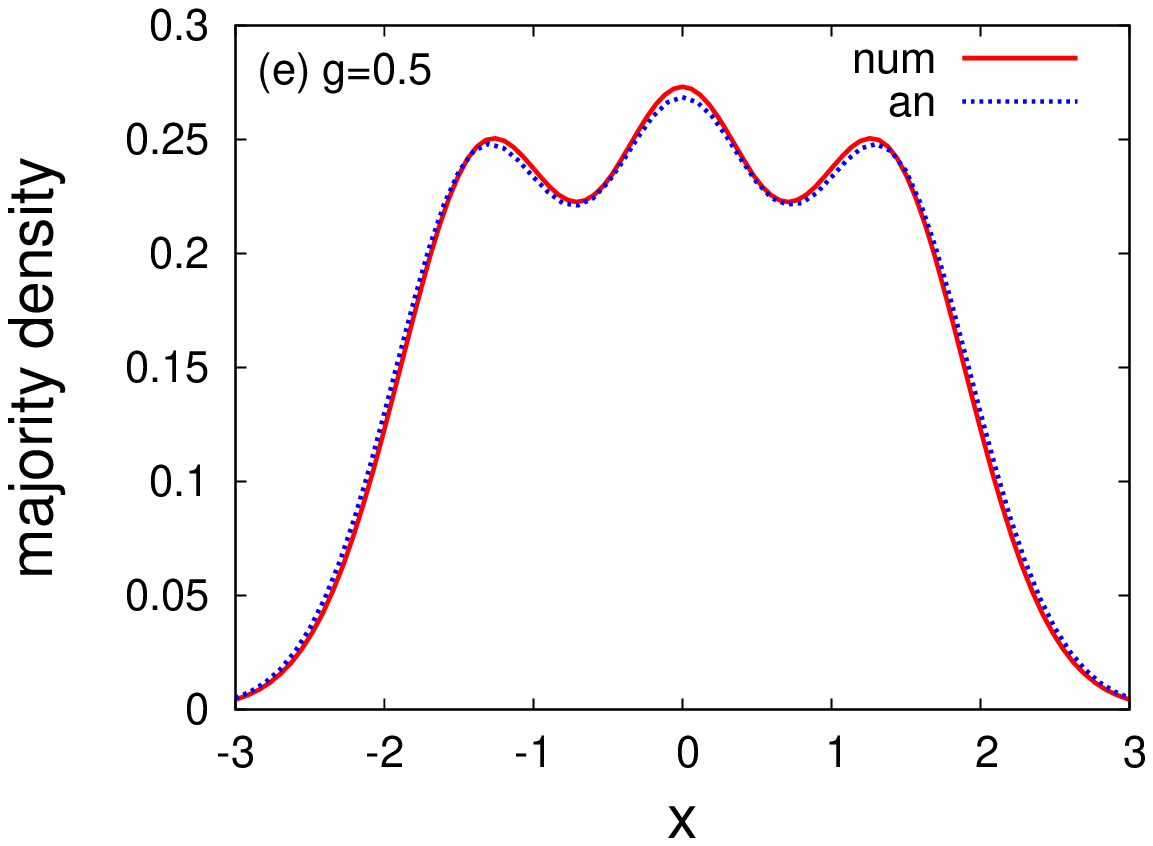}
\includegraphics[width=6.0 cm,height=6.0 cm]{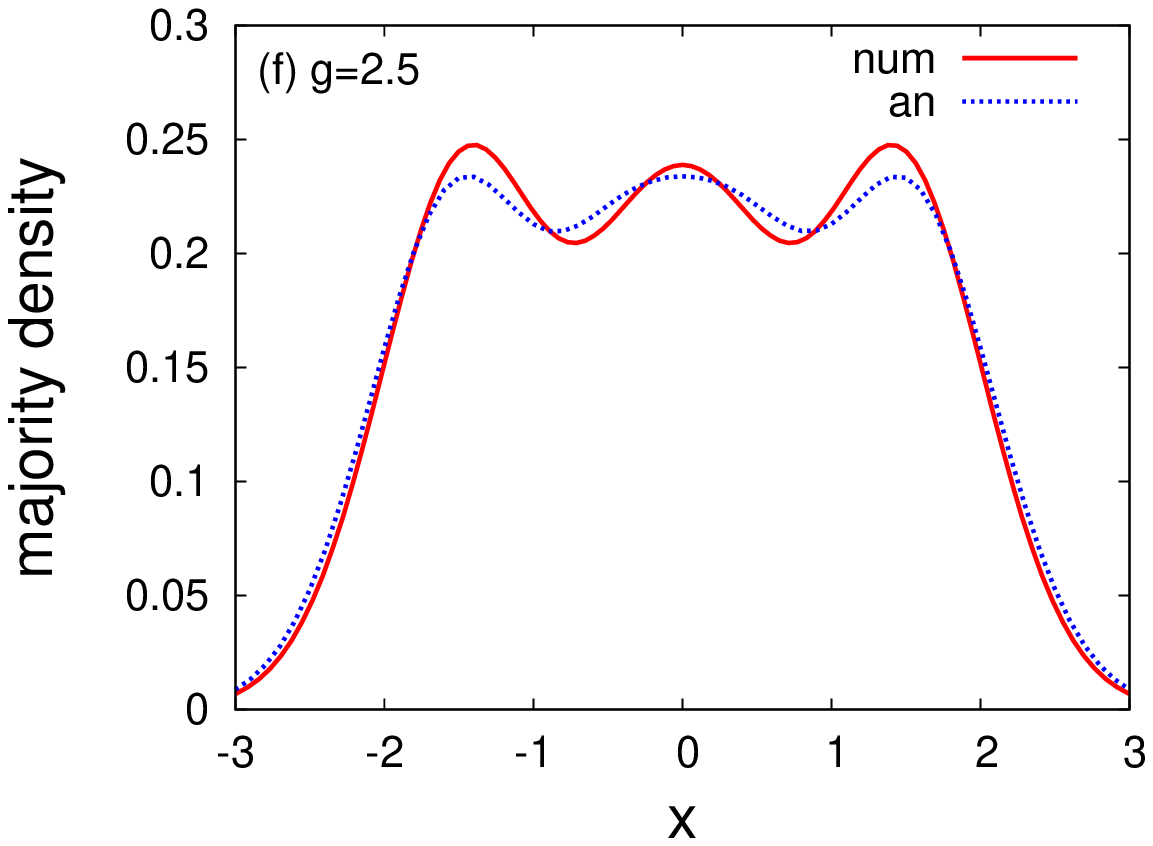}
\includegraphics[width=6.0 cm,height=6.0 cm]{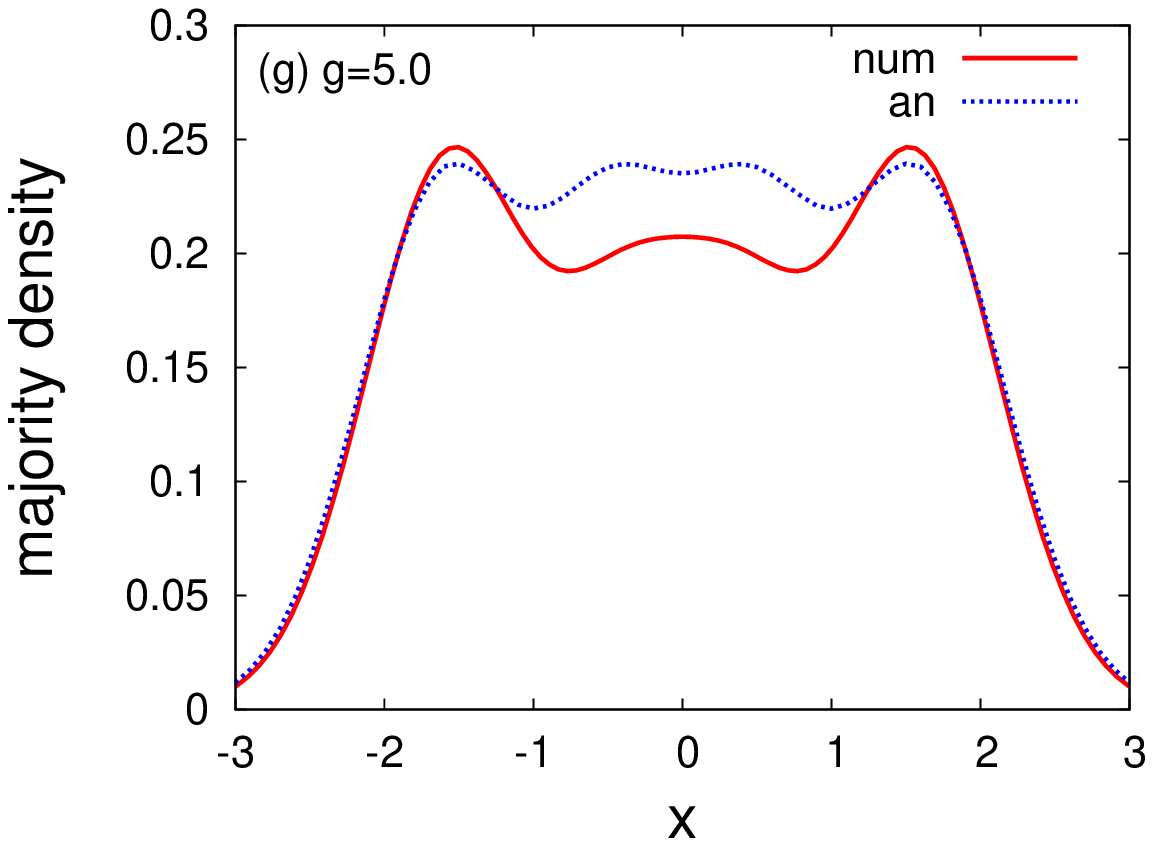}
\includegraphics[width=6.0 cm,height=6.0 cm]{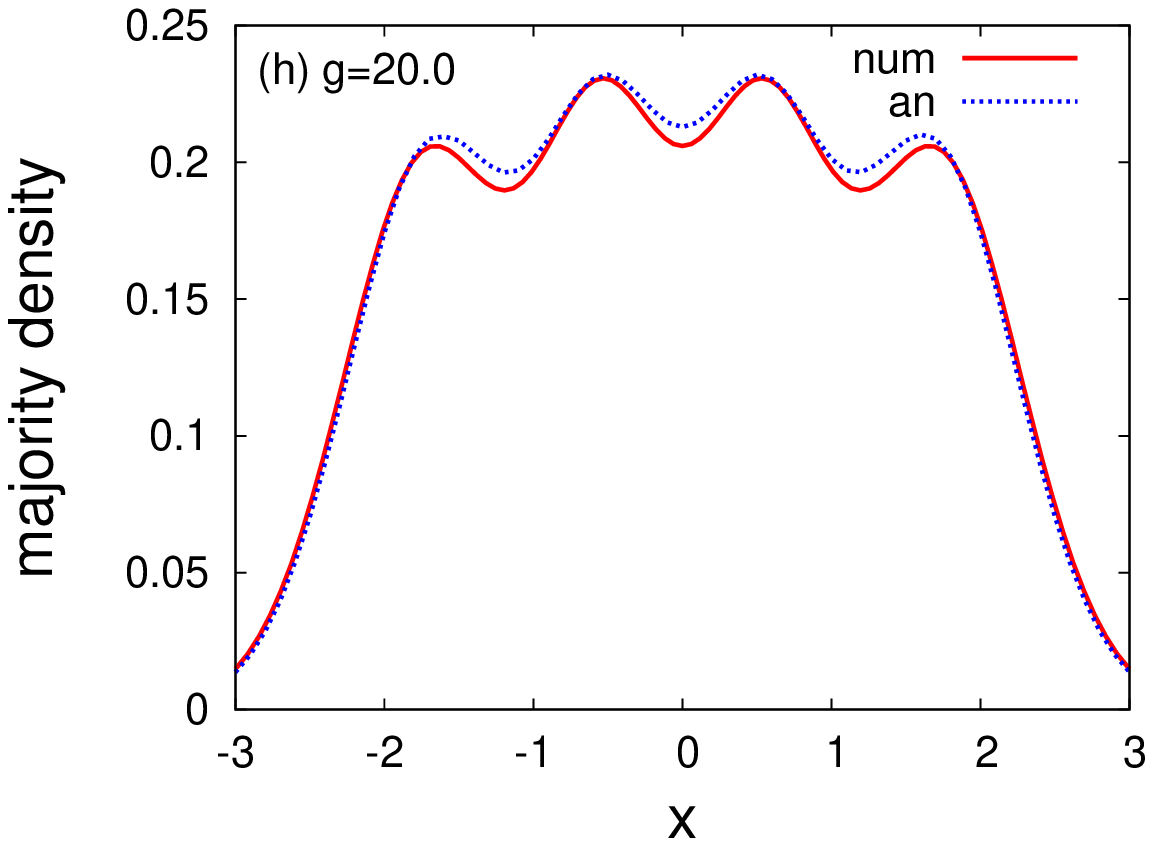}
\caption{One body densities of (a-d) the impurity  and  (e-h) majority fermions for the 3:1 case with varying interaction strength $g$ using the numerical (MCTDH) approach, and the analytical (CPWF) Ansatz.}
\label{fig3}
\end{figure*}

\section{Equal populations and partial imbalance}

In the following we analyze the density profiles of two cases which exhibit a different crossover to fermionization than the impurity physics discussed above. Since the Ansatz of the CPWF possesses a similar accuracy as already depicted in Fig. 1 (a) we will concentrate here on the physical effects for these cases and present only the numerically obtained densities in Fig. 4. 

Fig. 4 (a) shows the one-body density profiles for different interaction strengths for the case of equal populations (2:2). Note that the two hyperfine states possess exactly the same profile features in this completely balanced case. The two initial peaks for weak $g=0.5$ get broadened and flattened ($g=2.0$) and start already for intermediate $g=5.0$ to acquire side peaks, resulting in 4 distinct peaks for a strong coupling $g=15$ according to the fermionization theorem. The crossover is here much smoother than in the case of the impurity (reminiscent of the bosonic case \cite{brouzos,sascha}) with each peak (corresponding to a pair) spiting to two. The behaviour for equal populations can be also generalized to higher particle numbers, and we expect no substantial differences which we have checked for the few body case of 3:3 where the 3 initial peaks evolve to 6 again due to repulsive pair splitting.

Another very interesting scenario is that of a partially imbalanced population, which we illustrate here for the case 3:2 in Fig. 4 (b) and (c)  showing the density profiles with increasing coupling strengths for the minority and the majority atoms, respectively. The two minority atoms possess  2 lobes for weak interactions $g=0.5$ which are slightly pushed outwards in the trap for intermediate interactions $g=5.0,10$ forming a small plateau in the middle. Only for a strong coupling $g=14$ a third maxima rises in the middle of the potential and some outermost side wings are forming, which result to two additional peaks at $g=16$. We see again here a rather abrupt and interesting behaviour for very strong interactions, like in the case of the impurity fermion. The majority atoms   [Fig. 4 (c)] also hold the three-lobe profile for intermediate  interactions ($g=0.5, 5.0, 10$), while from the plateaus forming between the peaks two additional peaks emerge near $g=14.0$ resulting in a profile with five maxima. Inspecting Figs.4 (b) and (c) the two peaks of the minority density ($g=10$) which are located at the positions of the two plateaus of the majority density seem to represent a rather stable configuration for the repulsive particles. Still the tendency to fermionize in order to avoid each other as much as possible breaks this configuration giving rise to the additional peaks.  

The above mechanisms of the crossover to fermionization are present also in ensembles with larger number of atoms, yet they are more prominent and illustrative for few-atom systems. The connection between these mechanisms and the many-body phases like BCS and FFLO, is an open research pathway.

\begin{figure}
\includegraphics[width=8.0 cm,height=8.0 cm]{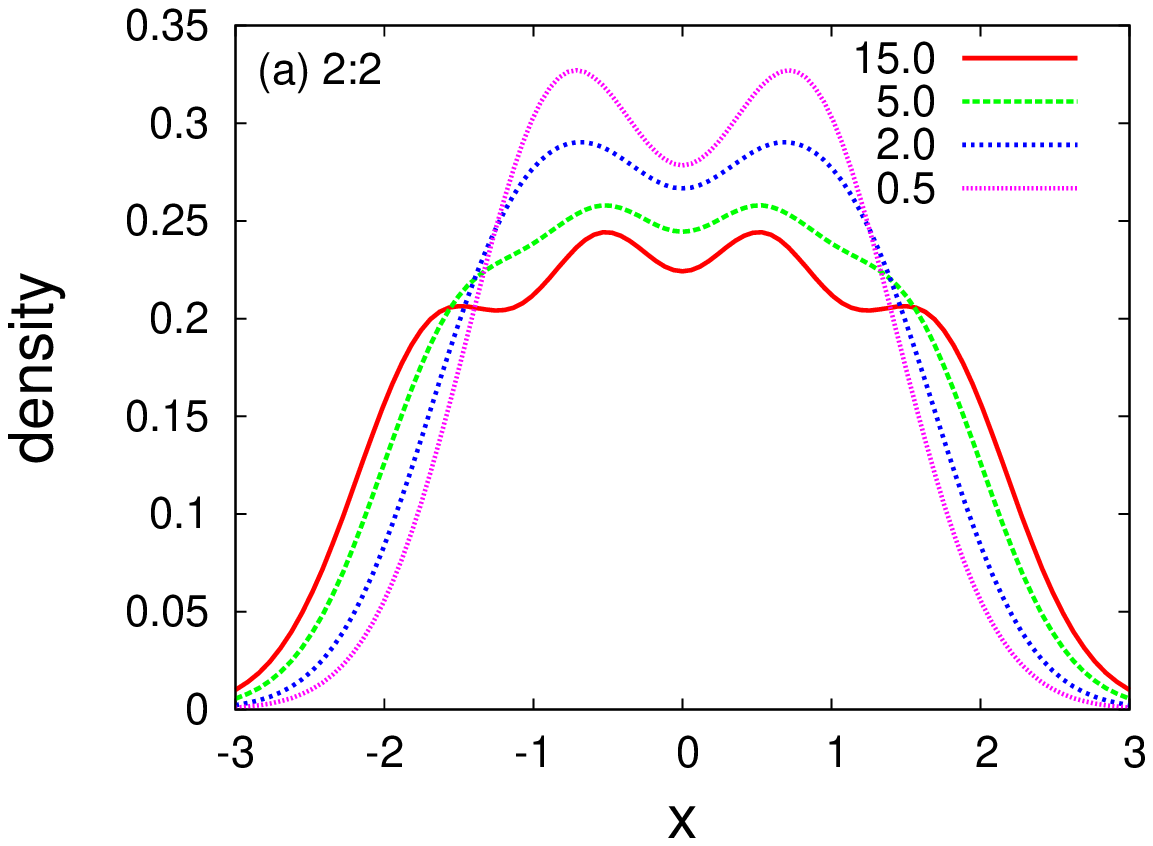} \\ 
\includegraphics[width=8.0 cm,height=8.0 cm]{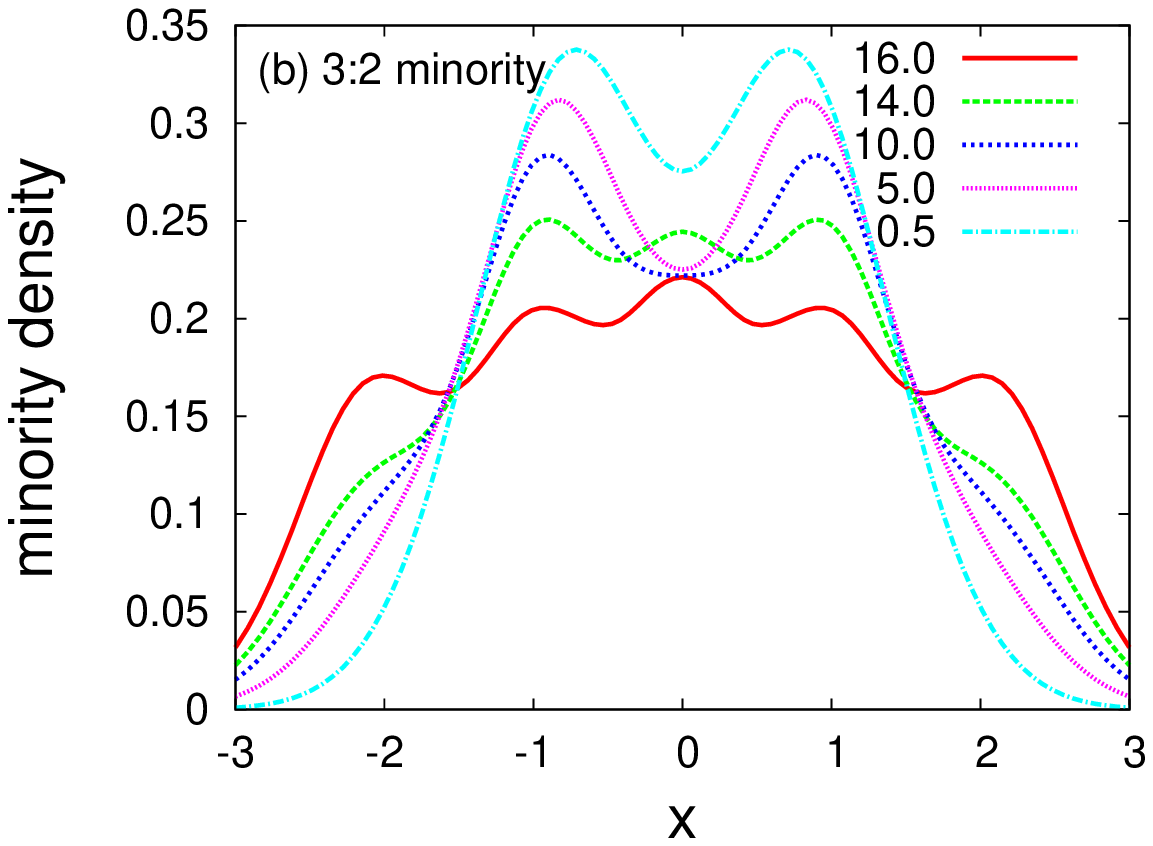}
\includegraphics[width=8.0 cm,height=8.0 cm]{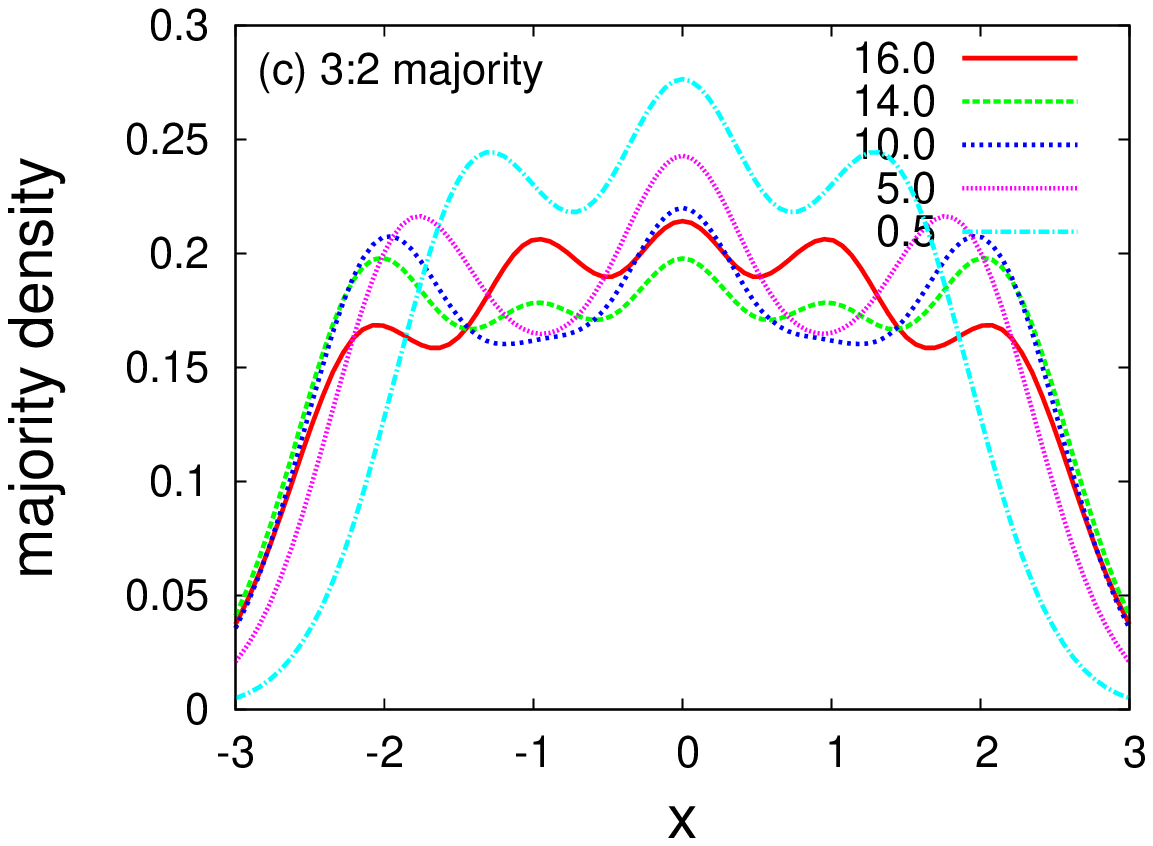}
\caption{one body densities for the case (a) 2:2 (the densities for the two components are identical) and for the case 3:2 for the (b) minority and (c) majority fermions.}
\label{fig4}
\end{figure}

\section{Concluding remarks and outlook}

We have employed exact numerical calculations (MCTDH) and an analytical approach (correlated-pair wave function CPWF) to explore the properties of few-fermion two-component mixtures in an one-dimensional harmonic trap. Our Ansatz for the function of the relative motion CPWF which is here generalized to cover the case of mixtures, is shown to be in a very good agreement (for the energies and the densities) with the numerical results of MCTDH (used here for the first time for the few-fermion case). We examined three general cases: (a) an impurity in a sea of majority fermions, where we have shown a strong variation of the density with an increasing strength of the repulsive interaction both for the impurity atom and the majority ones, (b) an equal population case for which the behaviour is smoother with two peaks arising in the place of one as the coupling strength increases for each pair of atoms and (c) a partially imbalanced case where very close to the resonance additional peaks arise on top of the plateaus in the density. 

Most of these theoretical considerations are also studied in  a very controllable manner in experiments of few fermion systems with a deterministically preparable atom number  \cite{selim,selim1,selim2}. The generalization of our Ansatz for the CPWF to mixtures is open to further improvements and applications which we initiated and sketched in this work, as well as to the extraction of other observables unraveling mechanisms and physical properties characterizing few-body systems in connection with and analogue to the corresponding many-body physics. 

\acknowledgments
I.B. is thankful to Andre Wenz, Gerhard Z\"urn, Thomas Lompe, and  Selim Jochim for stimulating discussions and their hospitality in Heidelberg.  

\appendix
\section{Computational Method}

In this work we have used  for the numerical calculations the Multi-Configurational Time-Dependent Hartree (MCTDH) method \cite{mctdhbook,meyer90,beck00}, which is a wave-packet dynamical tool with an outstanding efficiency for high-dimensional systems. 
The method has been applied  successfully to study few-boson dynamics and their stationary properties \cite{sascha}. Here we apply this method for first time to fermionic systems. Later developments of MCTDH include by construction the bosonic (MCTDHB) \cite{mctdhb} or the fermionic (MCTDHF) \cite{mctdhf} character of the system. We resort here to the initial Heidelberg MCTDH package \cite{mctdh:package} implemented for distinguishable particles and enforce the correct antisymmetry by the corresponding choice of the coefficients of the MCTDH expansion. 

The underlying idea of MCTDH is to treat the time-dependent Schr\"odinger equation
\begin{equation}
\left\{ \begin{array}{c}
i\dot{\Psi}=H\Psi\\
\Psi(Q,0)=\Psi_{0}(Q)\end{array}\right.\label{eq:TDSE}
\end{equation}
as an initial-value problem by an expansion in terms of Hartree products $\Phi_{J}$:
\begin{eqnarray}
\nonumber
\Psi(Q,t)&=&\sum_{J}A_{J}(t)\Phi_{J}(Q,t)
\nonumber\\
&\equiv&\sum_{j_{1}=1}^{n_{1}}\ldots\sum_{j_{f}=1}^{n_{f}}A_{j_{1}\ldots j_{f}}(t)\prod_{\kappa=1}^{f}\varphi_{j_{\kappa}}^{(\kappa)}(x_{\kappa},t),\label{eq:mctdh-ansatz}
\end{eqnarray}
where $J=(j_{1}\dots j_{f})$ are the different configurations and $f=N$ denotes the number of degrees of freedom and $Q\equiv(x_{1},\dots,x_{f})^{T}$. The single-particle functions SPFs $\varphi_{j_{\kappa}}^{(\kappa)}$ are in turn represented in a fixed, primitive basis implemented on a grid. For indistinguishable particles as in our case within the same component of fermions, the single-particle functions for each degree of freedom are chosen to be the first $N_M$ or $N_m$ harmonic oscillator states (identical for each atom belonging to the same component).

In the above expansion, both the coefficients $A_{J}$ and the Hartree products $\Phi_{J}$ are time-dependent. Using the Dirac-Frenkel variational principle, one can derive equations of motion for both $A_{J},\Phi_{J}$. This offers a huge advantage since the basis $\{\Phi_{J}(\cdot,t)\}$ is variationally optimal at each time $t$, and thus we can choose a small number of SPFs. The correct permutation symmetry which is important here can be enforced on the coefficients $A_{J}$ such that the total expansion is antisymmetric with respect to permutation of two atoms of the same species.  

In addition the Heidelberg MCTDH package  \cite{mctdh:package} incorporates the so-called \emph{relaxation method} which provides a way to obtain the lowest eigenstates of the system by propagating some wave function $\Psi_{0}$ by the non-unitary $e^{-H\tau}$ propagation in imaginary time. As $\tau\to\infty$, this automatically damps out any contribution but that stemming from the true ground state, \[ e^{-H\tau}\Psi_{0}=\sum_{J}e^{-E_{J}\tau}|J\rangle\langle J|\Psi_{0}\rangle.\] In our case it is important that the ground state is not as in the bosonic case given for free as the lowest state of indistinguishable particles system, but we have to apply a more sophisticated scheme termed improved relaxation \cite{meyer03} where one starts for $g=0$ with the two-component fermionic ground state placing the fermions on the Fermi ladder correctly. Here $\langle\Psi|H-E|\Psi\rangle$ is minimised with respect to both the coefficients $A_{J}$ and the configurations $\Phi_{J}$. The equations of motion are solved iteratively, first for $A_{J}(t)$ (by diagonalisation of $\langle\Phi_{J}|H|\Phi_{K}\rangle$ with fixed $\Phi_{J}$) and then propagating $\Phi_{J}$ in imaginary time over a short period. The cycle will then be repeated until convergence is achieved.

The computational effort of this method scales exponentially with the number of degrees of freedom, $n^{N}$. Just as an illustration, using $15$ orbitals and $N=5$ requires $7.6\cdot10^{5}$ configurations $J$. This restricts our analysis in the current setup to 5 atoms totally, given that we need to cover the whole range of interactions and especially the stronger coupling regime.  By contrast, the dependence on the primitive basis, and thus on the grid points, is not as severe. In our case, the grid spacing should of course be small enough to sample the interaction potential. Since the truncation of the Hilbert space (to $15$ or $20$ orbitals for $N=4,5$ respectively) is unavoidable one may employ additional tricks that heal this problem, as the rescaling of the interaction parameter proposed in \cite{dieter}. The 1D systems with the known fermionization limit allow for such a treatment that we employ here as explained in more detail in Section III.

\end{document}